\documentclass[twocolumn, notitlepage, aps, pra, superscriptaddress]{revtex4-1}

\usepackage{graphicx}
\usepackage{amsmath,amssymb,mathtools}
\usepackage{color}
\usepackage[colorlinks=true,linkcolor=blue,citecolor=blue]{hyperref}

\makeatletter
\def\l@subsection#1#2{}
\def\l@subsubsection#1#2{}
\makeatother

\newcommand{\rev}[1]{\textcolor{black}{#1}}
\newcommand{\revsecond}[1]{\textcolor{black}{#1}}

\begin{document}

\title{Theory of coupled parametric oscillators beyond coupled Ising spins}

\author{Marcello Calvanese Strinati}
\affiliation{Department of Physics, Bar-Ilan University, 52900 Ramat-Gan, Israel}
\author{Leon Bello}
\affiliation{Department of Physics and BINA Center of Nanotechnology, Bar-Ilan University, 52900 Ramat-Gan, Israel}
\author{Avi Pe'er}
\affiliation{Department of Physics and BINA Center of Nanotechnology, Bar-Ilan University, 52900 Ramat-Gan, Israel}
\author{Emanuele G. Dalla Torre}
\affiliation{Department of Physics, Bar-Ilan University, 52900 Ramat-Gan, Israel}

\date{\today}

\begin{abstract}
Periodically driven parametric oscillators offer a convenient way to simulate classical Ising spins. When many parametric oscillators are coupled dissipatively, they can be analogous to networks of Ising spins, forming an effective coherent Ising machine (CIM) that efficiently solves computationally hard optimization problems. In the companion paper, we studied experimentally the minimal realization of a CIM, i.e. two coupled parametric oscillators [L. Bello, M. Calvanese Strinati, E. G. Dalla Torre, and A. Pe'er, Phys. Rev. Lett. \textbf{123}, 083901 (2019)]. We found that the presence of an energy-conserving coupling between the oscillators can dramatically change the dynamics, leading to everlasting beats, which transcend the Ising description. Here, we analyze this effect theoretically by solving numerically and, when possible, analytically the equations of motion of two parametric oscillators. Our main tools include: (i) a Floquet analysis of the linear equations, (ii) a multi-scale analysis based on a separation of time scales between the parametric oscillations and the beats, and (iii) the numerical identification of limit cycles and attractors. Using these tools, we fully determine the phase boundaries and critical exponents of the model, as a function of the intensity and the phase of the coupling and of the pump. Our study highlights the universal character of the phase diagram and its independence on the specific type of nonlinearity present in the system. Furthermore, we identify new phases of the model with more than two attractors, possibly describing a larger spin algebra.
\end{abstract}


\maketitle

\tableofcontents

\section{Introduction}
Parametric oscillations are one of the best known examples of nontrivial effect induced by a periodic drive. Over the past decades, parametric oscillators have attracted a significant attention thanks to their wide range of applications for electronic low-noise amplification~\cite{PhysRev.106.384,PhysRev.107.317,4065451,5243812,doi:10.1063/1.1734980}. In recent years, parametric oscillators are used as generators of squeezed light~\cite{PhysRevA.29.408,PhysRevA.30.1386,JOSAB.4.001465, Lvovsky_2015}, with applications in high-accuracy sensing~\cite{PhysRevD.23.1693,Harry2010,Aasi2013,qdm}, quantum information and communication~\cite{Furusawa1998,Ralph1999,PhysRevA.61.010303,Braunstein2000,s41377-018-0011-3,s41467-018-03083-5}. They have also been studied in the context of nano- or microelectromechanical systems (NEMS or MEMS) both for practical applications and because they represent a suitable platform to analyse fundamental aspects of nonlinear dynamics~\cite{PhysRevB.67.134302,inbookronlifshitz,PhysRevE.79.026203,PhysRevE.80.046202,PhysRevE.84.016212,PhysRevLett.106.094102,PhysRevLett.108.264102,PhysRevE.94.022216}.

A \rev{degenerate} parametric oscillator is the canonical example of a period doubling instability. Due to the external periodic pump, the system can display two regimes: a stable regime, in which the oscillator is not excited, and a parametrically amplified regime, in which the oscillator oscillates at half the frequency of the pump. In the latter case, the equation of motion of the parametric oscillator admits two solutions, characterised by a relative shift of one period of the pump. A parametric oscillator is therefore the simplest example of a discrete time crystal that explicitly breaks time-translational symmetry~\cite{PhysRevLett.109.160401,PhysRevLett.116.250401,PhysRevLett.117.090402,PhysRevB.94.085112,PhysRevLett.118.030401,choi2017observation,sacha2017time,yao2018classical,sullivan2018dissipative,doi:10.1063/PT.3.4020,PhysRevLett.122.015701}. As such, it is suitable for the simulation of a classical spin-1/2 (Ising spin) where the spin states (`up' or `down') are represented by the two inequivalent solutions.

Recently, networks of many coupled \rev{degenerate} optical parametric oscillators have been proposed as suitable platforms to simulate networks of classical spin-1/2 (Ising) systems on a large scale. This kind of simulator was referred to as coherent Ising machine (CIM)~\cite{PhysRevA.88.063853}, and was proposed as a new platform to efficiently solve complex combinatorial and minimization problems. Because of its potential applications in computation, its recent experimental realization~\cite{nphoton.2016.68,s41534-017-0048-9,takesue2018large2dising} has triggered a significant amount of work, both on the theoretical and computational side~\cite{king2018emulating,Hamerly:18,1805.05217,CerveraLierta2018exactisingmodel,Tiunov:19,arXiv:1903.07163}.

In light of such potential applications, in this work, we focus on the minimal realization of such network, namely, \emph{two} coupled \rev{degenerate} parametric oscillators. Such system has been analyzed in previous studies in the context of MEMS~\cite{PhysRevE.84.016212,PhysRevLett.106.094102,PhysRevLett.108.264102}. In the companion letter~\cite{PhysRevLett.123.083901}, we analyzed the system of two coupled parametric oscillators experimentally and theoretically, in view of its application as the building block for a CIM. The two parametric oscillators were experimentally implemented by two radio-frequency cavities, in the presence of a power-splitter non-dissipative coupling. Our main finding was that the system of two coupled oscillators presents a much richer phenomenology than previously analyzed, depending on the values of the system parameters. In addition to a phase where the two oscillators display the expected behaviour of two Ising spins~\cite{PhysRevA.88.063853}, it was found that, depending on the dissipative or non-dissipative nature of the coupling, the system gives rise to a phase where the dynamics is characterized by limit cycles. In this case, the amplitudes of the two oscillators exhibit periodic beats on top of the fast oscillations at half the frequency of the drive. Such phase lies beyond the dynamics of coupled Ising spins, and its presence may be either a useful resource for CIMs or an additional source of error, which requires further investigation. The goal of this paper is to provide a theoretical background for the experiment in Ref.~\cite{PhysRevLett.123.083901}, as well as a complete characterization of the possible phases in the system of two coupled parametric oscillators.

Specifically, we first consider in Sec.~\ref{sec:linearparametricoscillatorbyfloquetansatz} the linearized equations of motion of the model and solve them using Floquet theory. This approach allows us to determine the stability diagram of the model. Next, in Sec.~\ref{sec:nonlinearcasebymultiplescaleanalysis}, we apply a multiple-scale analysis to determine the phase diagram in the presence of non-linearities. In our study we focus on the universal properties of the model, such as the critical exponents of the different instabilities and the role of the different types of non-linearities (Sec.~\ref{sec:discussionsondifferentmodels}). While most of the article is dedicated to energy preserving couplings, towards the end (Sec.~\ref{sec:dissipativecouplngandcim}) we consider a dissipative coupling and connect our findings to the results of Ref.~\cite{PhysRevA.88.063853} \rev{in the context of CIMs}.

\section{Linear parametric oscillators - Solution by Floquet theorem}
\label{sec:linearparametricoscillatorbyfloquetansatz}
We open our analysis by introducing a model of coupled parametric oscillators with purely energy-preserving coupling and showing the explicit solution in the absence of nonlinear terms. In our study, we do not follow the canonical approach to this problem, used for instance in Ref.~\cite{landau1982mechanics} to solve a single parametric oscillator. Here, we instead rely on the Floquet theorem (see e.g. Refs.~\cite{magnus1979hill,chicone2006ordinary,1367-2630-17-9-093039}), which can be easily generalized to more complicated situations.

\subsection{In-phase pumps}
\label{sec:inphasepumps}
In the absence of nonlinearities, a pair of coupled parametric oscillators is described by a set of two generalized linear Mathieu's equations:
\begin{equation}
\begin{array}{l}
\ddot x_1+\omega^2_0[1+h\,\sin(\gamma t)]x_1+\omega_0g\,\dot x_1-\omega_0r\,\dot x_2=0\vspace{0.1cm}\\
\ddot x_2+\omega^2_0[1+h\,\sin(\gamma t)]x_2+\omega_0g\,\dot x_2+\omega_0r\,\dot x_1=0
\end{array}\,\, .
\label{eq:conventionallyadoptednotationequationsofmotion1}
\end{equation}
In Eq.~\eqref{eq:conventionallyadoptednotationequationsofmotion1}, $\omega_0$ denotes the proper frequency of the oscillators, $h$ and $\gamma$ represent the intensity and frequency of the pumps, respectively, $g$ is the intrinsic loss term, which we take equal for both oscillators. The coupling $r$ describes an energy preserving coupling between the oscillators: this coupling corresponds to rotations in the $(x_1,x_2)$ plane and preserves the total energy, which is proportional to $x^2_{1,2}$ for the $x_1$ or $x_2$ modes, respectively. In the experiment of Ref.~\cite{PhysRevLett.123.083901} this coupling was implemented by a power splitter coupler. \rev{In Eq.~\eqref{eq:conventionallyadoptednotationequationsofmotion1}, the two oscillators are coupled such that the exchange of energy from $x_1$ to $x_2$, and \textit{vice versa}, occurs with the same rate, which is determined by $r$. This assumption will be relaxed in Sec.~\ref{sec:dissipativecouplngandcim}.} In the limit of $r\rightarrow0$. Eq.~\eqref{eq:conventionallyadoptednotationequationsofmotion1} becomes equivalent to two decoupled parametric oscillators described by two Mathieu's equations.

The equations in Eq.~\eqref{eq:conventionallyadoptednotationequationsofmotion1} can be separated by performing the change of basis $x_{\pm}(t)=x_1(t)\pm i\,x_2(t)$. In such basis, we have two decoupled parametric oscillators with real and imaginary loss terms:
\begin{subequations}
\begin{align}
&\ddot x_++\omega^2_0[1+h\,\sin(\gamma t)]x_++\omega_0(g+i\,r)\dot x_+=0\\
&\ddot x_-+\omega^2_0[1+h\,\sin(\gamma t)]x_-+\omega_0(g-i\,r)\dot x_-=0 \,\, .
\end{align}
\label{eq:conventionallyadoptednotationequationsofmotion2}
\end{subequations}
The $\dot x_{\pm}$ terms in Eq.~\eqref{eq:conventionallyadoptednotationequationsofmotion2} can be reabsorbed into the definitions of the fields by introducing $x_{\pm}(t)=e^{-(g\pm i\,r)\omega_0t/2}\,y_\pm(t)$ and then Eq.~\eqref{eq:conventionallyadoptednotationequationsofmotion2} becomes
\begin{equation}
\ddot y_\pm+\omega^2_0\left[1-\frac{{(g\pm i\,r)}^2}{4}+h\,\sin(\gamma t)\right]y_\pm=0 \,\, .
\end{equation}
For simplicity, we first focus on the case of $g,r\ll1$, where one can neglect terms proportional to $(g\pm ir)^2$:
\begin{equation}
\ddot y_\pm+\omega^2_0\left[1+h\,\sin(\gamma t)\right]y_\pm=0 \,\, .
\vspace{0.2cm}
\label{eq:conventionallyadoptednotationequationsofmotion3}
\end{equation}
In this limit, $y_+(t)$ and $y_-(t)$ obey the same equation. In the next subsection we will show how to release this constraint. Since the equations of motion are periodic with period $\mathcal{T}=2\pi/\gamma$, we can look for solutions of the form $y_\pm(t)=e^{-i\mu t}\,f(t)$, where $\mu$ is a complex frequency and $f(t)$ is a periodic function with period $\mathcal{T}$. In order to determine $\mu$, we can proceed as follows: using the periodicity of $f(t)$, we can express $y_\pm(t)$ in terms of its Fourier components:
\begin{equation}
y_\pm(t)=e^{-i\mu t}\sum_{n\,\in\,\mathbb{Z}}A_n\,e^{in\gamma t} \,\, ,
\label{eq:conventionallyadoptednotationequationsofmotion4}
\end{equation}
where $A_n$ identifies the amplitude of the $n$-th Fourier component, where $n$ is an integer number. If we plug Eq.~\eqref{eq:conventionallyadoptednotationequationsofmotion4} into Eq.~\eqref{eq:conventionallyadoptednotationequationsofmotion3} and equate to zero terms multiplying the same oscillating factor, we obtain a recursive equation for the coefficients $A_n$:
\begin{equation}
D_n(\mu)\,A_n+\cfrac{i}{2}\,\omega^2_0h\left(A_{n+1}-A_{n-1}\right)=0 \,\, ,
\label{eq:parametricresonanceancondition}
\end{equation}
where we define $D_n(\mu)=\omega^2_0-{(n\gamma-\mu)}^2$. We can more conveniently write Eq.~\eqref{eq:parametricresonanceancondition} in the matrix form
\begin{widetext}
\begin{equation}
\left(\begin{array}{ccccccc}
\ddots& & &\vdots & & & \reflectbox{$\ddots$} \\
 & -i\,\cfrac{\omega^2_0h}{2} & D_{-1}(\mu) & i\,\cfrac{\omega^2_0h}{2} & 0& 0 & \\\\
\cdots& 0 & -i\,\cfrac{\omega^2_0h}{2} & D_{0}(\mu) & i\,\cfrac{\omega^2_0h}{2} & 0 &\cdots \\\\
& 0 & 0 & -i\,\cfrac{\omega^2_0h}{2} & D_{+1}(\mu) & i\,\cfrac{\omega^2_0h}{2} & \\
\reflectbox{$\ddots$}& & &\vdots & & & \ddots
\end{array}\right)\left(
\begin{array}{c}\vdots \\A_{-1}\\\\\\A_0\\\\\\A_{+1}\\\vdots\end{array}
\right)=0 \,\, ,
\label{eq:matrixrelationforanparametricoscillator0}
\end{equation}
\end{widetext}
The expression in Eq.~\eqref{eq:matrixrelationforanparametricoscillator0} can be rewritten as $\mathbf{M}_\mu\cdot\mathbf{A}=0$, where $\mathbf{A}$ identifies the column vector containing the Fourier components of $f(t)$, and $\mathbf{M}_\mu$ is the infinite-by-infinite matrix in Eq.~\eqref{eq:matrixrelationforanparametricoscillator0}. In general, from Eq.~\eqref{eq:matrixrelationforanparametricoscillator0}, the requirement for the existence of a nontrivial solution requires that the determinant of the matrix $\mathbf{M}_\mu$ vanishes, ${\rm det}(\mathbf{M}_\mu)=0$.

We now make the following observation: as we see from Eq.~\eqref{eq:parametricresonanceancondition}, the parametric drive directly couples each Fourier component $A_n$ only to its nearest-neighbour ones $A_{n\pm1}$. This means that $A_n$ is coupled to $A_{n\pm m}$, with $m>0$, via a coupling that is of the order of $h^{m}$. Within a perturbative fashion, at first order in $h$, we therefore see that the strongest effect of the parametric drive is coupling $A_n$ with $A_{n-1}$ or with $A_{n+1}$. By inspection of Eq.~\eqref{eq:parametricresonanceancondition}, we see that, for $h\rightarrow0^+$, such coupling occurs when $D_n(\mu)=D_{n\pm1}(\mu)=0$. This condition is satisfied for $\mu=\omega_0(2n\pm1)$, which corresponds to the parametric resonance condition $\gamma=2\omega_0$.

Since the function in Eq.~\eqref{eq:conventionallyadoptednotationequationsofmotion4} is periodic with period $\gamma=2\omega_0$, we can without loss of generality focus on the case of $n=0$ and therefore consider the situation in which only $A_0$ and $A_{-1}$ are coupled by the parametric drive. Therefore, from Eq.~\eqref{eq:parametricresonanceancondition} and Eq.~\eqref{eq:matrixrelationforanparametricoscillator0} the requirement for the existence of a nontrivial solution reduces to the condition
\begin{equation}
\hspace{-1cm}{\rm det}\left(
\begin{array}{cc}
\omega^2_0-{(\mu+\gamma)}^2 & i\,\cfrac{\omega^2_0h}{2}\\
-i\,\cfrac{\omega^2_0h}{2} & \omega^2_0-\mu^2
\end{array}
\right)\equiv P_4(\mu)=0 \,\, ,
\label{eq:parametricresonanceancondition2}
\end{equation}
with $\gamma$ sufficiently close to $2\omega_0$. We can parametrize the deviation from the parametric resonance condition by introducing a small detuning $\epsilon$ and by rewriting $\gamma=2\omega_0+\epsilon$. The polynomial $P_4(\mu)$ defined in Eq.~\eqref{eq:parametricresonanceancondition2} has four complex roots. Since we are looking for the parametric resonance between $A_0$ and $A_{-1}$, we consider only the degenerate roots that converge to $\mu=-\omega_0$ when $h=0$ and $\epsilon=0$, which are found to be
\begin{eqnarray}
&&\hspace{-0.4cm}\mu_{\pm}(h,\epsilon)=-\omega_0-\frac{\epsilon}{2}\nonumber\\
&&\hspace{-0.4cm}\pm\frac{1}{2}\sqrt{\epsilon^2+4\epsilon\omega_0+8\omega^2_0-2\omega_0\sqrt{4{(\epsilon+2\omega_0)}^2+\omega_0^2h^2}} \,\, .
\label{eq:parametricoscillatorrootsfloquet}
\end{eqnarray}
As evident from Eq.~\eqref{eq:conventionallyadoptednotationequationsofmotion4}, if $\mu_{1,2}$ are complex, then their nonzero imaginary part quantifies the rate of the parametric exponential damping [${\rm Im}(\mu)<0$] or exponential amplification [${\rm Im}(\mu)>0$] of the solution $y_{\pm}(t)$. By expanding $\mu_{1,2}$ for small $h$ and $\epsilon$ and by discarding terms of the order of $\epsilon^2h$, the two solutions in Eq.~\eqref{eq:parametricoscillatorrootsfloquet} can be written as
\begin{equation}
\mu_\pm(h,\epsilon)\simeq-\omega_0-\frac{\epsilon}{2}\pm\frac{i}{2}\sqrt{{\left(\frac{\omega_0h}{2}\right)}^2-\epsilon^2} \,\, .
\label{eq:mu1mu2expressionfornonzeroepsilon}
\end{equation}
The region such that ${\rm Im}(\mu_\pm)>0$ (i.e., $\omega_0 h/2 > |\epsilon|$) identifies the \emph{\rev{linear} instability region}, where parametric amplification occurs. Within this region, as evident from Eq.~\eqref{eq:conventionallyadoptednotationequationsofmotion4} and Eq.~\eqref{eq:mu1mu2expressionfornonzeroepsilon}, the oscillator frequency is always exactly equal to $\omega_0+\epsilon/2\equiv\gamma/2$, i.e., half of the pump frequency, therefore manifesting its time-crystal nature. Thus, the growing solution of Eq.~\eqref{eq:conventionallyadoptednotationequationsofmotion3} is found to be $y_+(t)\sim e^{(\omega_0t/2)\sqrt{(h/2)^2-(\epsilon/\omega_0)^2}}\,\cos(\gamma t/2)$, from which it follows that
\begin{equation}
x_\pm(t)\sim e^{\mp i\,\frac{\omega_0rt}{2}}\,e^{\frac{\omega_0t}{2}\left(\sqrt{{\left(\frac{h}{2}\right)}^2-\left(\frac{\epsilon}{\omega_0}\right)^2}-g\right)}\cos(\gamma t/2) \,\, .
\label{eq:parametricoscillatorsolutionfloquet}
\end{equation}
Equation~\eqref{eq:parametricoscillatorsolutionfloquet} describes two solutions $x_1(t)={\rm Re}[x_\pm(t)]$ and $x_2(t)=\pm{\rm Im}[x_\pm(t)]$. Within the instability region, these solutions represent two parametrically-driven solutions with in-quadrature beats (in the limit $r\ll1$). Notice that, from Eq.~\eqref{eq:parametricoscillatorsolutionfloquet}, it is evident that the parametric amplification occurs when the pump strength $h$ is above a threshold value: in this case, parametric amplification occurs when $\sqrt{(h\omega_0/2)^2-\epsilon^2}>\omega_0g$. At resonance $\epsilon=0$ (i.e., $\gamma=2\omega_0$), the threshold condition reads $h>2g$.

\subsection{Varying the pump phase - General derivation}
\label{sec:dephasedpumps}
We now use the scheme introduced in Sec.~\ref{sec:inphasepumps} in order to solve the system of linear Mathieu's equations in the case when the two oscillators are pumped with a different phase, $\gamma t \to \gamma t \pm \phi/2$:
\begin{equation}
\begin{array}{l}
\ddot x_1+\omega^2_0\left[1+h\,\sin(\gamma t-\phi/2)\right]x_1+\omega_0g\,\dot x_1-\omega_0r\,\dot x_2=0\vspace{0.1cm}\\
\ddot x_2+\omega^2_0\left[1+h\,\sin(\gamma t+\phi/2)\right]x_2+\omega_0g\,\dot x_2+\omega_0r\,\dot x_1=0 \,\, ,
\end{array}
\label{eq:coupledparametricoscillatorequationslinearphi2}
\end{equation}
where $\phi$ represents the phase difference between the two pumps. Since the oscillators normally lock to the phase of the pump, varying the phase difference between the pumps is equivalent to varying the phase between the two oscillators. We can equivalently rewrite Eq.~\eqref{eq:coupledparametricoscillatorequationslinearphi2} as
\begin{equation}
\begin{array}{l}
\ddot x_1+\omega^2_0\left[1+h\,\sin(\gamma t)\cos(\phi/2)\right]x_1\vspace{0.1cm}\\
-h\omega_0^2\sin(\phi/2)\,\cos(\gamma t)\,x_1+\omega_0g\,\dot x_1-\omega_0r\,\dot x_2=0\vspace{0.1cm}\\
\ddot x_2+\omega^2_0\left[1+h\,\sin(\gamma t)\cos(\phi/2)\right]x_2 \vspace{0.1cm}\\
+h\omega_0^2\sin(\phi/2)\,\cos(\gamma t)\,x_2+\omega_0g\,\dot x_2+\omega_0r\,\dot x_1=0 \,\, .
\end{array}
\label{eq:coupledparametricoscillatorequationslinearphi3}
\end{equation}
In the basis $x_\pm(t)=x_1(t)\pm i\,x_2(t)$, Eq.~\eqref{eq:coupledparametricoscillatorequationslinearphi3} becomes
\begin{equation}
\begin{array}{l}
\ddot x_++\omega^2_0\left[1+h\,\sin(\gamma t)\cos(\phi/2)\right]x_+\vspace{0.1cm}\\
+\omega_0(g+i\,r)\dot x_+-h\omega_0^2\sin(\phi/2)\,\cos(\gamma t)\,x_-=0\vspace{0.1cm}\\
\ddot x_-+\omega^2_0\left[1+h\,\sin(\gamma t)\cos(\phi/2)\right]x_-\vspace{0.1cm}\\
+\omega_0(g-i\,r)\dot x_--h\omega_0^2\sin(\phi/2)\,\cos(\gamma t)\,x_+=0 \,\, .
\end{array}
\label{eq:coupledparametricoscillatorequationslinearphi4}
\end{equation}
By comparing Eq.~\eqref{eq:coupledparametricoscillatorequationslinearphi4} with Eq.~\eqref{eq:conventionallyadoptednotationequationsofmotion2}, one finds that a finite dephasing $\phi\neq0$ has two effects: (i) it reduces the strength of the parametric drive from $h$ to $h\cos(\phi/2)$ and (ii) gives birth to an effective coupling between the $x_+(t)$ and $x_-(t)$ oscillators, whose strength is proportional to $\sin(\phi/2)$. Since now the $x_\pm(t)$ oscillators obey different equations of motion, we write the Floquet form of $x_\pm(t)$ [see Eq.~\eqref{eq:conventionallyadoptednotationequationsofmotion4}] in the vector form:
\begin{equation}
\left(\begin{array}{c}x_+\\x_-\end{array}\right)=e^{-i\mu t}\sum_{n\,\in\,\mathbb{Z}}\,e^{in\gamma t}\,\left(\begin{array}{c}H^{(+)}_n\\H^{(-)}_n\end{array}\right) \,\, ,
\label{eq:conventionallyadoptednotationequationsofmotion4phi}
\end{equation}
where $H^{(+)}_n$ and $H^{(-)}_n$ represent the $n$-th Fourier component of $x_+$ and $x_-$, respectively. By proceeding as done for Eq.~\eqref{eq:parametricresonanceancondition}, we obtain the recursion relations for $H^{(\pm)}_n$:
\begin{eqnarray}
&&D_{\pm,n}(\mu)\,H^{(\pm)}_n+i\,\frac{\omega_0h^2}{2}\,\cos\left(\frac{\phi}{2}\right)\left(H^{(\pm)}_{n+1}-H^{(\pm)}_{n-1}\right)\nonumber\\
&&\hspace{1cm}-\frac{\omega_0h^2}{2}\,\sin\left(\frac{\phi}{2}\right)\left(H^{(\mp)}_{n-1}+H^{(\mp)}_{n+1}\right)=0 \,\, ,
\label{eq:equationforhplusandhminusparametricoscillatorfloquet1}
\end{eqnarray}
where we define $D_{\pm,n}=\omega^2_0-{(n\gamma-\mu)}^2+i\,\omega_0(g\pm i\,r)(n\gamma-\mu)$. As done for Eq.~\eqref{eq:matrixrelationforanparametricoscillator0}, Eq.~\eqref{eq:equationforhplusandhminusparametricoscillatorfloquet1} can be also written in the matrix form
\begin{widetext}
\begin{equation}
\left(\begin{array}{ccccccc}
\ddots& & &\vdots & & & \reflectbox{$\ddots$} \\
 & \mathbf{M}^*(h,\phi) & \mathbf{D}_{-1}(\mu) & \mathbf{M}(h,\phi) & 0 & 0 & \\\\
\cdots & 0 & \mathbf{M}^*(h,\phi) & \mathbf{D}_{0}(\mu) & \mathbf{M}(h,\phi) & 0 &\cdots \\\\
& 0 & 0 & \mathbf{M}^*(h,\phi) & \mathbf{D}_{+1}(\mu) & \mathbf{M}(h,\phi) & \\
\reflectbox{$\ddots$}& & &\vdots & & & \ddots
\end{array}\right)\left(
\begin{array}{c}\vdots \\\mathbf{H}_{-1}\\\\\\\mathbf{H}_0\\\\\\\mathbf{H}_{+1}\\\vdots\end{array}
\right)=0 \,\, ,
\label{eq:matrixrelationforanparametricoscillator0phi}
\end{equation}
\end{widetext}
where one defines
\begin{subequations}
\begin{align}
&\hspace{-0.1cm}\mathbf{D}_{n}(\mu)=\left(\begin{array}{cc}D_{+,n}&0\\0&D_{-,n}\end{array}\right) \\
&\hspace{-0.1cm}\mathbf{M}(h,\phi)=\left(\begin{array}{cc}i\,\cfrac{h\omega^2_0}{2}\,\cos(\phi/2)&-\cfrac{h\omega^2_0}{2}\,\sin(\phi/2)\\\\-\cfrac{h\omega^2_0}{2}\,\sin(\phi/2)&i\,\cfrac{h\omega^2_0}{2}\,\cos(\phi/2)\end{array}\right) \,.
\label{eq:matrixdandmparametricoscillatorphi}
\end{align}
\end{subequations}
We now define $\Omega_{r}=\omega_0\sqrt{1+(r^2-g^2)/4}$ and focus on the limit of small $g$, where
\begin{equation}
D_{\pm,n}\simeq\Omega^2_{r}-{\left[n\gamma-\mu-\frac{\omega_0}{2}\,(ig\pm r)\right]}^2 \,\, .\\
\label{eq:expressionfordplusminus}
\end{equation}
As in Sec.~\ref{sec:inphasepumps}, we focus on the four-by-four minor of the matrix in Eq.~\eqref{eq:matrixrelationforanparametricoscillator0phi} which contains the blocks for $n=0$ and $n=-1$. As for Eq.~\eqref{eq:parametricresonanceancondition2}, the requirement for the existence of a nontrivial solution results into the computation of the roots of an (eight-order) polynomial, $P_8(\mu)$. Among its eight roots, some of  them (which we call the relevant roots) can acquire a nonzero imaginary part depending on which parametric resonance is met. As we will explain below, in this case, there are three distinct resonances, where parametric amplification can occur, which are given by $\gamma=2\Omega_r,2\Omega_r\pm\omega_0r$.

\subsubsection{Parametric resonance at $\gamma=2\Omega_r$}
We first discuss the parametric resonance at $\gamma=2\Omega_r$. In the limit of $h=0$, one has \emph{four} relevant roots that are doubly degenerate. They are $\mu^{(0)}_{\pm,\pm}\simeq\pm\omega_0r/2-\gamma/2-i\,\omega_0g/2$, for which $D_{\pm,0} = D_{\pm,-1}=0$. For finite $h$, the root degeneracy is removed and one finds
\begin{equation}
\mu^{(0)}_{\pm,\pm}(h,\gamma)\simeq\pm\frac{\omega_0r}{2}-\frac{\gamma}{2}-i\,\frac{\omega_0g}{2}\pm i\,\Gamma^{(0)}(h,\gamma) \,\, .
\label{eq:eigenvaluesparametricoscillatorfloquetphi}
\end{equation}
As for the case in Eq.~\eqref{eq:mu1mu2expressionfornonzeroepsilon}, the additional imaginary part $\Gamma^{(0)}(h,\gamma)$, determines the rate of parametric amplification for the resonance around $\gamma=2\Omega_r$. The rate $\Gamma^{(0)}$ is identical for $x_+$ and $x_-$, indicating that both quadratures are amplified by the same amount. Importantly, the oscillation frequencies of the two quadratures differ by $\omega_0 r$, indicating that the system's energy oscillates between $x_1$ and $x_2$. Also, $x_{\pm}(t)$ display fast oscillations at half the frequency of the pump but with in-quadrature beats at frequency $\omega_0r/2$ on top of such oscillations. As we will discuss in Sec.~\ref{sec:nonlinearcasebymultiplescaleanalysis} (see also Appendix~\ref{sec:oscilaltionthresholdandbeats}), in the presence of nonlinearities, this behaviour will evolve into a limit cycle that will eventually stabilize the amplitude of the beats.

\subsubsection{Parametric resonances at $\gamma=2\Omega_r\pm\omega_0r$}
We now move to the resonances at $\gamma=2\Omega_r\pm\omega_0r$. In this case, we have only \emph{two} relevant roots, which are degenerate in the limit $h\rightarrow0^+$. We denote the two roots by $\mu^{(\pm\omega_0r)}_{\pm}\simeq-\gamma/2-i\,\omega_0g/2$. A nonzero $h$ removes the degeneracy in the two relevant roots, adding an equal and opposite imaginary part, which now we call $\Gamma^{(\pm\omega_0r)}(h,\gamma,r)$. Therefore, for finite $h$, one has
\begin{equation}
\mu^{(\pm\omega_0r)}_{\pm}(h,\gamma)\simeq-\frac{\gamma}{2}-i\,\frac{\omega_0g}{2}\pm i\,\Gamma^{(\pm\omega_0r)}(h,\gamma,r)\,\, .
\label{eq:eigenvaluesparametricoscillatorfloquetphi2}
\end{equation}
From Eq.~\eqref{eq:eigenvaluesparametricoscillatorfloquetphi2}, we see that both oscillators oscillate with a frequency equal to $\gamma/2$ within all the instability region (for small $h$). Therefore, the system behaves as a time crystal, in which fast oscillations of $x_\pm$ are locked in phase with the even or odd cycles of the pump.

The functions $\Gamma^{(0)}(h,\gamma)$ and $\Gamma^{(\pm\omega_0r)}(h,\gamma,r)$ in Eqs.~\eqref{eq:eigenvaluesparametricoscillatorfloquetphi} and~\eqref{eq:eigenvaluesparametricoscillatorfloquetphi2} identify the three different instability regions around the three parametric resonances $\gamma=2\Omega_r,2\Omega_r\pm\omega_0r$, respectively. In contrast to the case discussed in Sec.~\ref{sec:inphasepumps}, we could not find an analytical expression for $\Gamma^{(0)}(h,\gamma)$ and $\Gamma^{(\pm\omega_0r)}(h,\gamma,r)$. Therefore, in order to determine the stability phase diagram, we resort to numerics.

\begin{figure}[t]
\centering
\includegraphics[width=8.7cm]{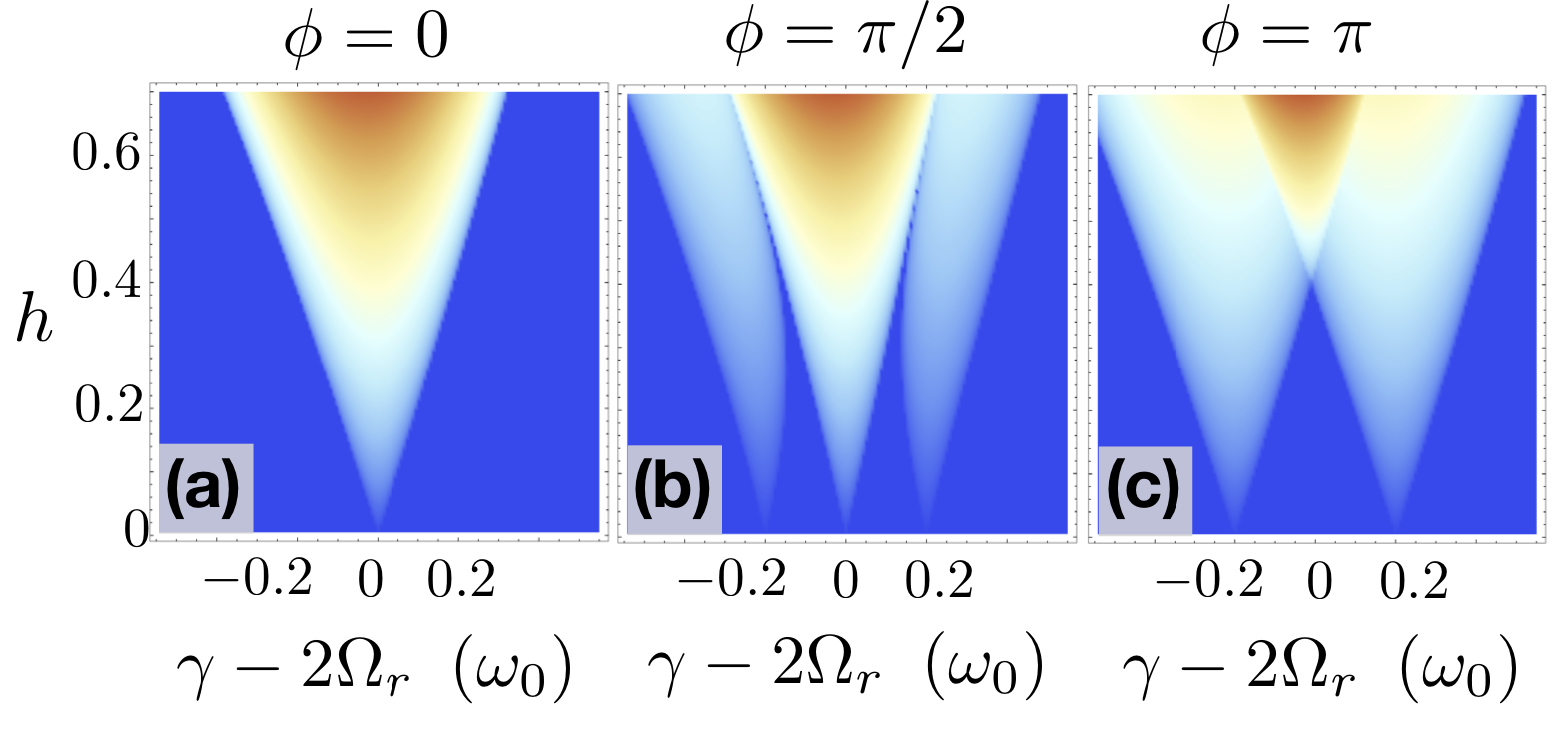}
\caption{Example of the instability region in the $h$ vs. $\gamma-2\Omega_r$ plane, in units of $\omega_0$. The instability regions, shown in arbitrary scale from zero (dark blue) to some maximum value (red), have been quantified by numerically computing the imaginary part of the roots in Eqs.~\eqref{eq:eigenvaluesparametricoscillatorfloquetphi} and~\eqref{eq:eigenvaluesparametricoscillatorfloquetphi2}, for small $g=10^{-2}$ and $r=0.2$. We use (a) $\phi=0$, (b) $\phi=\pi/2$ and (c) $\phi=\pi$. There can be a nonzero region in which different instability regions overlap (evident for $\phi=\pi$).}
\label{fig:instabilityregions}
\end{figure}

\begin{figure}[t]
\centering
\includegraphics[width=8.6cm]{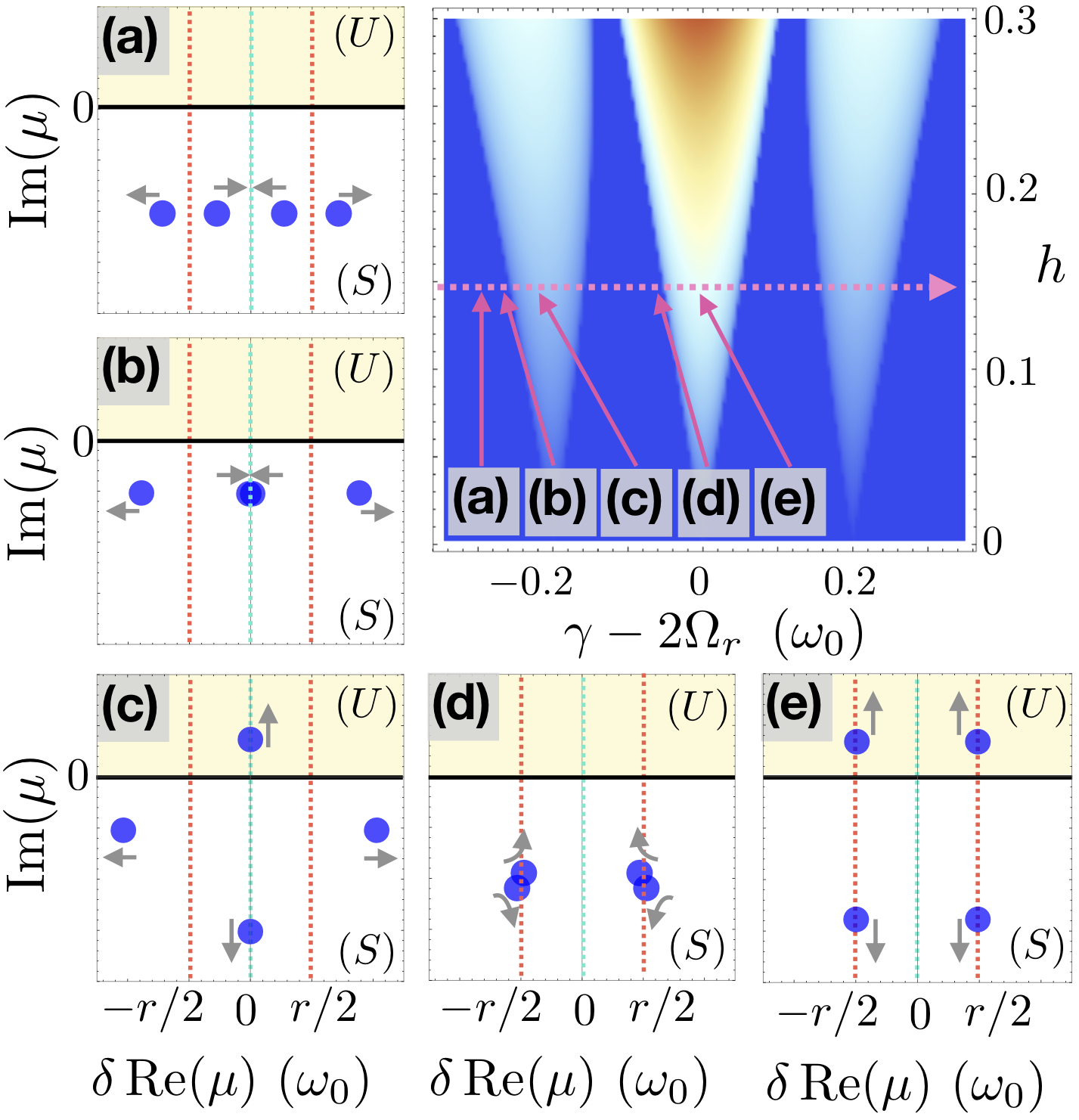}
\caption{Stability phase diagram as in Fig.~\ref{fig:instabilityregions} for $\phi=\pi/2$ (here reported up to $h=0.3$) and configurations of the four relevant roots [blue dots in panels (a)-(e)] as the phase diagram is horizontally cut at fixed $h$ (magenta dashed arrow) from left to right, and grey arrows indicate how the roots move accordingly. In the insets, we show the roots by plotting their imaginary part (in arbitrary units) versus their real part to which we subtract $\gamma/2$: $\delta\,{\rm Re}(\mu)={\rm Re}(\mu)-\gamma/2$, in units of $\omega_0$. The yellow area (U) denotes the unstable region [${\rm Im}(\mu)>0$], whereas the stable region (S) corresponds to ${\rm Im}(\mu)<0$.}
\label{fig:instabilityregions2}
\end{figure}

\subsection{\rev{Linear} instability regions}
\label{sec:instabilityregions}
The \rev{regions of linear instability} can be numerically determined from the imaginary part of the roots of the polynomial $P_8(\mu)$, for different values of $h$ and $\gamma$. This determines $\revsecond{-\omega_0g/2}+\Gamma^{(0)}(h,\gamma)$ or $\revsecond{-\omega_0g/2}+\Gamma^{(\pm\omega_0r)}(h,\gamma,r)$ in Eqs.~\eqref{eq:eigenvaluesparametricoscillatorfloquetphi} and~\eqref{eq:eigenvaluesparametricoscillatorfloquetphi2}. An example of the instability regions in the $h$ vs. $\gamma-2\Omega_r$ plane is shown in Fig.~\ref{fig:instabilityregions}. For concreteness, we show the instability regions for small $g$, which we choose $g=10^{-2}$, and $r=0.2$, and for (a) $\phi=0$, (b) $\phi=\pi/2$ and (c) $\phi=\pi$. For $\phi=0$, the instability region consists of one cone centred at $\gamma=2\Omega_r$. For nonzero $\phi$, two additional outer instability regions appear centred at $\gamma-2\Omega_r=\pm\omega_0r=\pm0.2\,\omega_0$ around the central region. For $\phi=\pi$, the resonance at $\gamma=2\Omega_r$ is completely suppressed, and only the instability regions around $\gamma-2\Omega_r=\pm\omega_0r$ are found.

A better insight regarding the properties of the system is given by studying the behaviour of the real and imaginary parts of the relevant roots as we vary $\gamma$ and $h$. This is shown in Fig.~\ref{fig:instabilityregions2}, focusing in particular on the instability phase diagram computed at $\phi=\pi/2$ [panel (b) of Fig.~\ref{fig:instabilityregions}], which is also reported in Fig.~\ref{fig:instabilityregions2} for completeness. The blue dots represent the four relevant roots [Eqs.~\eqref{eq:eigenvaluesparametricoscillatorfloquetphi} and~\eqref{eq:eigenvaluesparametricoscillatorfloquetphi2}], plotted by showing their imaginary parts as a function of their real parts from which we subtract $\gamma/2$. We do not show the other four roots since they do not contribute to the instabilities of the system and therefore are not relevant for the present discussion.

\begin{figure}[t]
\centering
\includegraphics[width=8cm]{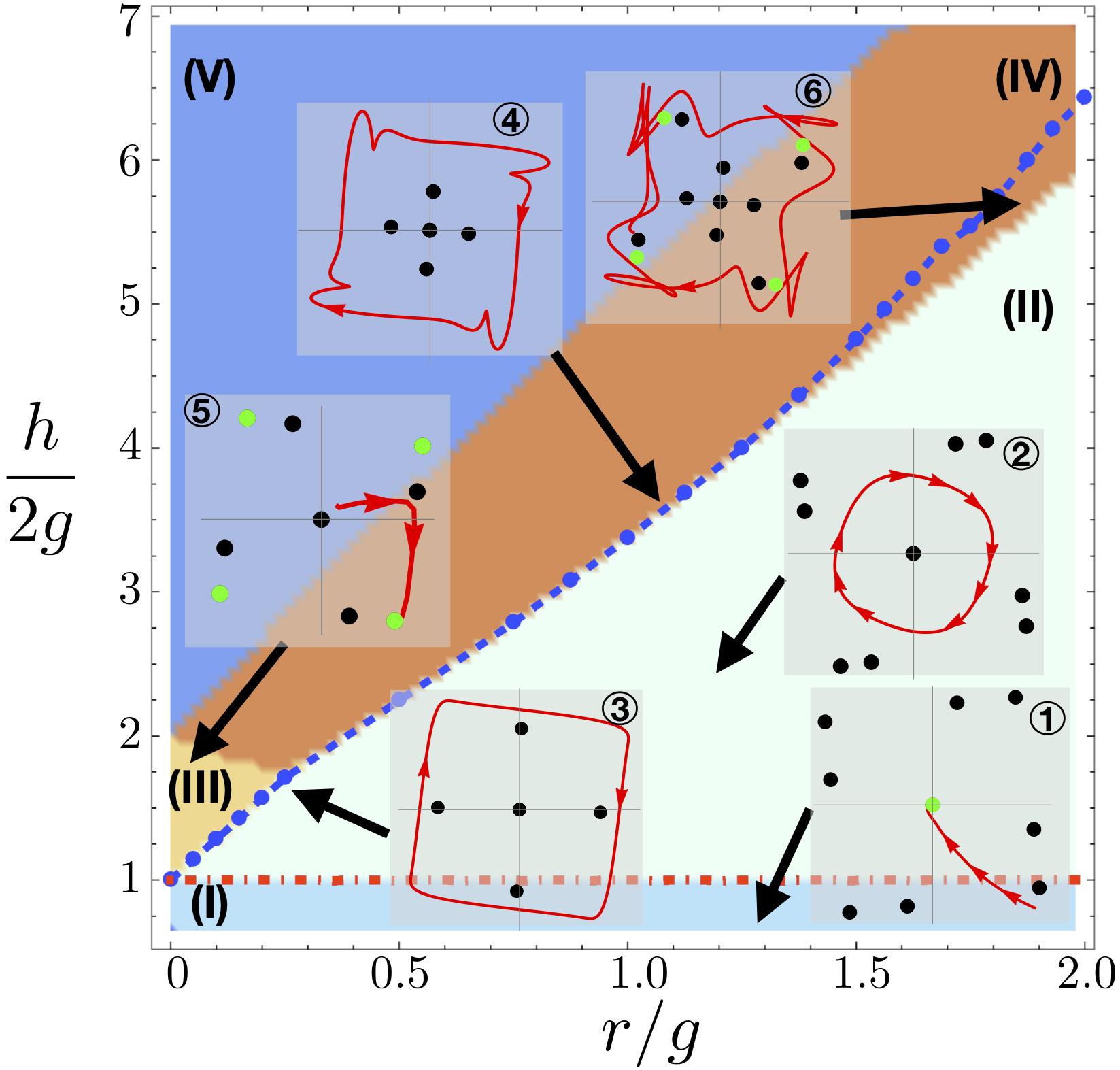}
\caption{Phase diagram of the nonlinear oscillators in the $h/(2g)$ vs. $r/g$ plane obtained by solving numerically Eq.~\eqref{eq:nonlinearmathieuequationflow2220coupledphihfixed} for $\phi=0$. Different phases correspond to different configurations of the fixed points. In the insets, we show the fixed points (black dots for saddle points and green dots for stable fixed points) and the flow (red line) in the $B_R$ vs. $A_R$ plane. For $h/(2g)$ not too far from the threshold $h/(2g)=1$, we can identify three main regions: (I) below the threshold, only the origin is a stable attractor (inset 1); (II) right above threshold, the origin loses its stability giving birth to a stable limit cycle, whose shape depends on the distance from the system threshold and on the coupling strength [inset 2 deep into region (II), insets 3 and 4 close to the boundary of region (II), for low and high pump power, respectively]. In this region, $x_\pm$ exhibits everlasting beats; (III) synchronization region, in which four stable attractors stabilize the dynamics (inset 5). In addition to region (III), we also find two additional regions (IV) and (V) in which eight or sixteen stable fixed points are found (not shown), respectively. Interestingly, between region (II) and region (IV), we find a subregion in which the limit cycle can coexists with stable attractors (inset 6). The blue dashed line and the red dash-dotted line enclose the region in which the limit cycle is found. The experimentally accessed region is usually up to $h/(2g)\simeq 2$. To the best of our knowledge, the regions that are found for larger pump intensities, in which the model displays a number of stable points larger than four, remains experimentally unexplored.}
\label{fig:phasediagramwithflowphi0}
\end{figure}

We show the relevant roots in five prototype cases in Fig.~\ref{fig:instabilityregions2}: (a) inside the stable region [${\rm Im}(\mu)<0$, i.e., the white area (S) of the insets], all roots have \revsecond{negative} imaginary part, which is equal to $\revsecond{-\omega_0g/2}$ in the small $g$ and $r$ limit. This case corresponds to an exponential damping in time for both oscillators. Inside the instability regions, one or two pairs of roots acquire in addition an equal and opposite imaginary part, depending on which one of the instability regions is entered. In the case of the outer instability regions [panel (c)], only two roots acquire an additional factor $\Gamma^{(\pm\omega_0r)}$ while having a real part that is locked to $\gamma/2$ [see also Eq.~\eqref{eq:eigenvaluesparametricoscillatorfloquetphi2}], which is highlighted in the figure by the cyan dashed vertical line. Instead, inside the central instability region [panel (e)], four roots acquire an additional imaginary part $\Gamma^{(0)}$ while having a real part locked to $\gamma/2\pm\omega_0r/2$ [see also Eq.~\eqref{eq:eigenvaluesparametricoscillatorfloquetphi}], which is instead highlighted by the red vertical lines. These two types of instabilities correspond, respectively, to a Pitchfork bifurcation and to a Hopf instability~\cite{strogatz2007nonlinear}. Parametric amplification occurs when the unstable region [${\rm Im}(\mu)>0$, i.e., the yellow area (U) in the insets] is entered, i.e., when the overall imaginary part is such that either $\revsecond{-\omega_0g/2}+\Gamma^{(\pm\omega_0r)}>0$ for the outer regions, or $\revsecond{-\omega_0g/2}+\Gamma^{(0)}>0$ for the central region, which identifies the threshold for parametric amplification.

The key result of this analysis is that, in our system, when the parametric \rev{linear} instability is met, the two modes are amplified and oscillate with a frequency that is locked to half of the frequency of the pump. The parametric amplification can occur with or without beats, depending on which \rev{region of linear instability} (the central or the outer ones, respectively) is entered. Such \rev{linearly unstable regions} are the precursors of the \rev{stable regions of} limit cycle and synchronized \rev{oscillations}, in which \rev{nonlinear effects eventually stabilize the long-time dynamics. This will be the topic of} the next sections (see also Appendix~\ref{sec:oscilaltionthresholdandbeats}).

As a final remark, we stress that the advantage of using the perturbative method here presented is that it grants us a good analytical control. Quantitatively, the result presented in this section are valid strictly speaking in the limit of $h\rightarrow0^+$. For a not too large finite value of $h$, there will be corrections to our findings, but the qualitative picture remains valid. For the sake of completeness, we mention that the full numerical solution can be obtained by resorting to the formalism of fundamental matrices~\cite{chicone2006ordinary}.

\section{Nonlinear case - Perturbative multiple-scale analysis}
\label{sec:nonlinearcasebymultiplescaleanalysis}
The method based on Floquet's theorem presented in Sec.~\ref{sec:linearparametricoscillatorbyfloquetansatz} allows us to systematically study systems of linear coupled parametric oscillators, but it cannot be applied in the presence of nonlinearities. When a nonlinear term is included in the equations of motion, one can resort to a multiple-scale perturbation method~\cite{kevorkian1996multiple} in order to determine the long-time dynamics of the oscillators. The goal of this section is to apply such method in order to study the dynamics of the system of coupled oscillators [Eq.~\eqref{eq:coupledparametricoscillatorequationslinearphi2}] in the specific case where a quadratic nonlinearity is included in the model.

\begin{figure}[t]
\includegraphics[width=7.9cm]{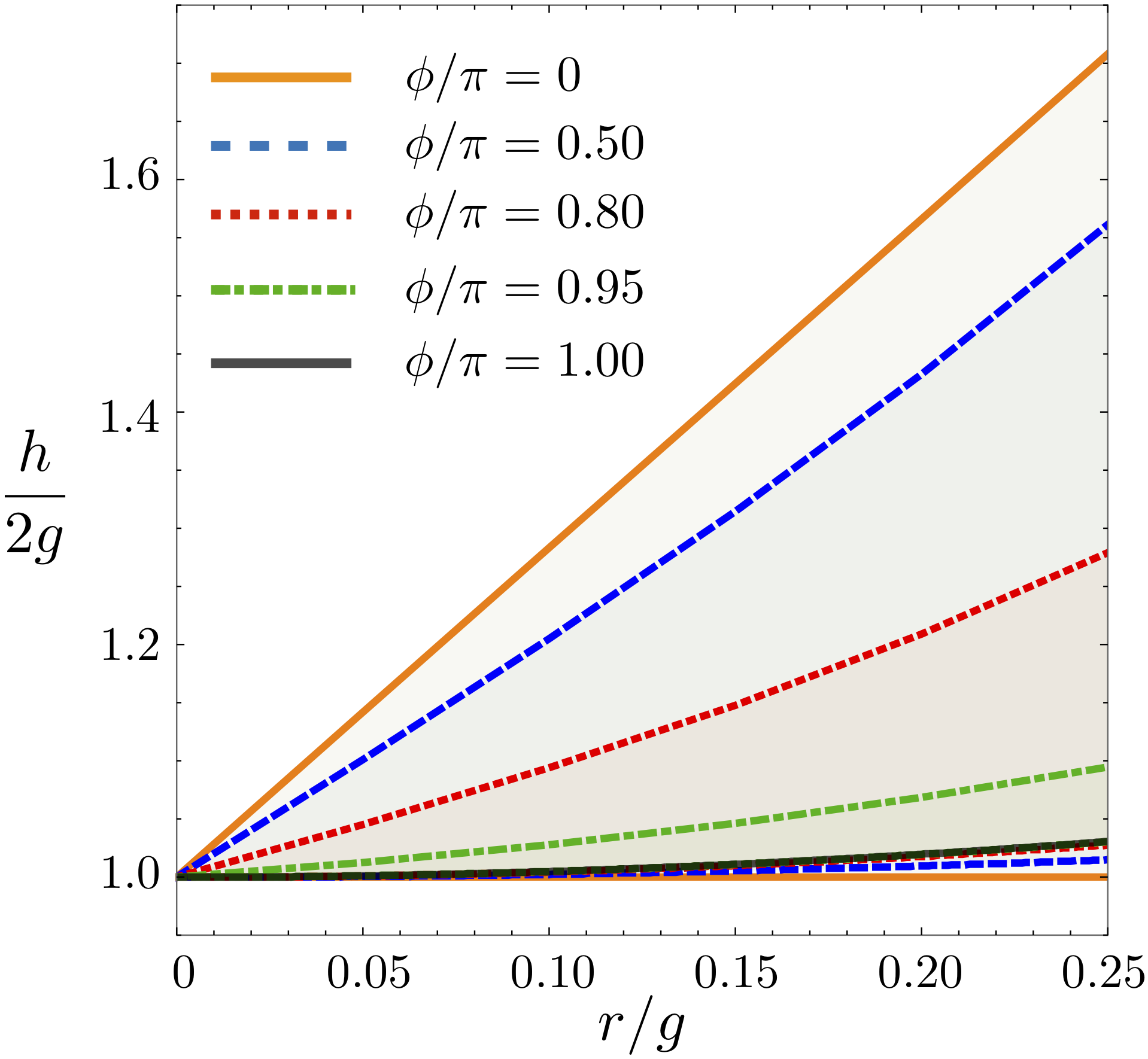}
\caption{Phase boundaries as in Fig.~\ref{fig:phasediagramwithflowphi0} for different values of the pump dephasing $\phi$. The shaded area enclosed by two curves determines the region in which the limit cycle is found, for a given value of $\phi$. In particular we show the boundaries for $\phi/\pi=0$ (full orange lines), $\phi/\pi=0.5$ (dashed blue lines), $\phi/\pi=0.8$ (dotted red lines), $\phi/\pi=0.95$ (dash-dotted green lines) and $\phi/\pi=1$ (black full line). The presence of $\phi\neq0$ lowers the boundary for the infinite-period bifurcation, and rises the boundary for the supercriticial Hopf bifurcation (see Appendix~\ref{sec:oscilaltionthresholdandbeats}).}
\label{fig:phasediagramdifferentphi}
\end{figure}

\subsection{Equations for the long-time dynamics}
\label{sec:coupledparametricoscillators}
In the actual physical context, there are different sources of nonlinearities that can appear in the equations of motion [Eq.~\eqref{eq:coupledparametricoscillatorequationslinearphi2}], such as saturation, Kerr effects or pump depletion. In this section, we focus on one type of nonlinearity, namely, the pump depletion, which is in many experimental contexts the most relevant type of nonlinearity. We postpone the discussion of other types of nonlinearities to Sec.~\ref{sec:discussionsondifferentmodels}.

The pump depletion accounts for the fact that the pump intensity is \revsecond{depleted within the nonlinear medium, by means of down-conversion processes to} the \revsecond{signal (and idler)} field. In the limit of small depletion, we can therefore write the equations of motion as
\begin{equation}
\begin{array}{l}
\ddot x_1+\omega^2_0\left[1+h\left(1-\beta\,x^2_1\right)\,\sin(\gamma t)\right]x_1\\\\
\hspace{3.5cm}+\omega_0g\,\dot x_1-\omega_0r\,\dot x_2=0\\\\
\vspace{-0.2cm}
\ddot x_2+\omega^2_0\left[1+h\left(1-\beta\,x^2_2\right)\,\sin(\gamma t+\phi)\right]x_2\\\\
\hspace{3.5cm}+\omega_0g\,\dot x_2+\omega_0r\,\dot x_1=0
\end{array} \,\, .
\label{eq:coupledparametricoscillatorequationswithnonlinearitydynamicalphigfixed}
\end{equation}
Here, $\beta$ quantifies the pump depletion. We now focus on the resonant case $\gamma=2\Omega_r$ where the system is more affected by the parametric instability. In order to determine the non-trivial long-time dynamics of the system, we proceed with a multiple-scale perturbative expansion~\cite{kevorkian1996multiple}. The details of the calculation are reported in Appendix~\ref{sec:detailsofhtederivation} for the sake of completeness.

In Eq.~\eqref{eq:coupledparametricoscillatorequationswithnonlinearitydynamicalphigfixed}, we identify $\omega_0$ as the largest frequency scale, which identifies the fastest time scale of the system $t=2\pi/\omega_0$. We assume that the coupling constants, $r$, $g$ and $h$ are much smaller than unity, and influence the dynamics of $x_1(t)$ and $x_2(t)$ only on time scales which are much longer than $2\pi/\omega_0$. In these conditions, the full dynamics can be separated in fast-varying and slow-varying degrees of freedom. If we work at fixed $g$, which we take as the small expansion parameter of the theory, we can identify the characteristic time scale of the slow-varying degrees of freedom by $\tau=gt$. Therefore, we can write $x_1(t,\tau)=A(\tau)\,e^{i\omega_0 t}+A^*(\tau)\,e^{-i\omega_0t}$ and $x_2(t,\tau)=B(\tau)\,e^{i\omega_0 t}+B^*(\tau)\,e^{-i\omega_0t}$, where $e^{\pm i\omega_0t}$ describes fast oscillations at frequency $\omega_0=\gamma/2$, and $A(\tau)$ and $B(\tau)$ represent the slow-varying complex amplitudes for $x_1$ and $x_2$, respectively. In the following, we express time in units of $\omega_0$, i.e., we define $\tilde\tau=\omega_0\tau$, and it is convenient to redefine $h$ and $r$ with respect to $g$, i.e., by introducing $\tilde h=h/g$ and $\tilde r=r/g$.

By separating the real and imaginary parts of each complex amplitude, i.e., $A=A_R+i\,A_I$ and $B=B_R+i\,B_I$, the main result is that the dynamics of the slow-varying amplitudes is described by a set of four coupled equations:
\begin{widetext}
\begin{subequations}
\begin{align}
&\frac{\partial A_R}{\partial\tilde\tau}=\left[\frac{\tilde h}{4}-\frac{1}{2}-\frac{\beta\tilde h}{2}\left(A^2_R+3\,A^2_I\right)\right]A_R+\frac{\tilde r}{2}\,B_R  \qquad \frac{\partial A_I}{\partial\tilde\tau}=\left[-\frac{\tilde h}{4}-\frac{1}{2}+\frac{\beta\tilde h}{2}\left(A^2_I+3\,A^2_R\right)\right]A_I+\frac{\tilde r}{2}\,B_I \label{eq:nonlinearmathieuequationflow2220coupledphihfixeda}\\
&\frac{\partial B_R}{\partial\tilde\tau}=\frac{\tilde h}{4}\left[B_R\,\cos(\phi)+B_I\,\sin(\phi)\right]-\frac{1}{2}\,B_R-\frac{\beta\tilde h}{2}\left[B^3_R\,\cos(\phi)+2\,B^3_I\sin(\phi)+3\,B_RB^2_I\,\cos(\phi)\right]-\frac{\tilde r}{2}\,A_R\\
&\frac{\partial B_I}{\partial\tilde\tau}=\frac{\tilde h}{4}\left[B_R\,\sin(\phi)-B_I\,\cos(\phi)\right]-\frac{1}{2}\,B_I-\frac{\beta\tilde h}{2}\left[2B^3_R\,\sin(\phi)-B^3_I\,\cos(\phi)-3\,B^2_RB_I\,\cos(\phi)\right]-\frac{\tilde r}{2}\,A_I \,\, .
\end{align}
\label{eq:nonlinearmathieuequationflow2220coupledphihfixed}
\end{subequations}
\end{widetext}

The system in Eq.~\eqref{eq:nonlinearmathieuequationflow2220coupledphihfixed} encodes the dynamics of $A$ and $B$. According to the standard analysis of nonlinear systems~\cite{strogatz2007nonlinear}, all the informations that we need in order to describe the long-time dynamics of $x_{1,2}(t)$ can be found by studying the configuration of the fixed points of Eq.~\eqref{eq:nonlinearmathieuequationflow2220coupledphihfixed}, which are found by imposing $\partial A/\partial\tilde\tau=\partial B/\partial\tilde\tau=0$. Their stability is determined by the eigenvalues of the Jacobian matrix at a specific point (see Appendix~\ref{sec:expressionforthejacobianmatrix}). Because we were not able to find an analytic expression for the fixed points, we resorted only to the numerical solution of Eq.~\eqref{eq:nonlinearmathieuequationflow2220coupledphihfixed}. For the configuration of the fixed points in the decoupled case ($\tilde r=0$) the reader is referred to Appendix~\ref{sec:decoupledparametricoscillators}

\subsection{Phase diagram of two coupled oscillators}
The key result for $\phi=0$ is reported in Fig.~\ref{fig:phasediagramwithflowphi0}, in which we show the phase diagram in the $h/(2g)$ vs. $r/g$ plane. According to our analysis, three main regions are found: when the system is below the threshold for parametric amplification [region (I)], the origin $A=B=0$ is the only stable attractor, and each trajectory of $A(\tau)$ and $B(\tau)$ is attracted into the origin. In this region, oscillations are suppressed in the long-time limit. 

For any finite $\tilde r$, as the pump intensity is increased up to a threshold value identified by the red dash-dotted line in Fig.~\ref{fig:phasediagramwithflowphi0}, the origin becomes a saddle point giving birth to a stable limit cycle in its surrounding via a supercritical Hopf bifurcation [region (II)]. In this region, the two oscillators display everlasting beats, whose frequency, close to the threshold, is determined by $\omega_0r$ and whose shape changes as $\tilde h$ is increased. For the analytical derivation of the boundary between region (I) and region (II) the reader is referred to Appendix~\ref{sec:oscilaltionthresholdandbeats}.

As the pump intensity is further increased, stable attractor and saddle nodes are born in pairs via saddle-node bifurcations. Specifically, depending on the value of $\tilde r$, by increasing $\tilde h$, a region with either four [region (III) ] or eight [region (IV)] stable fixed points is entered. \rev{Inside these regions}, the limit cycle disappears and the amplitude of the oscillations becomes constant in time (synchronized). The transition line between the two regimes has been numerically determined and is identified by the blue dashed line in the figure. We find that such a transition can occur in two different ways. For small couplings, we find that the period of the limit cycle, \rev{which represents the period of the beats on top of the fast oscillations at half the pump frequency}, diverges as region (III) is approached (see also Sec.~\ref{sec:criticalscalingbythreescaleanalysis}). At the boundary between region (II) and region (III), eight fixed points (four attractors and four saddle points) are born on the limit cycle via a saddle-node bifurcation, causing the extinction of the limit cycle as region (III) is entered. This phenomenology is customary referred to as an infinite-period bifurcation. For larger values of the coupling, we find that there is a first area inside region (IV), in which the limit cycle can coexist with the stable attractors and therefore fast oscillations occur either displaying beats or with constant amplitude, depending on the initial conditions. After this region, the limit cycle collapses into one of the attractors, and therefore only synchronized oscillations are found. For even larger values of $\tilde h$, a region with sixteen stable fixed points [region (V)] is found.

As $\phi$ is increased from $0$ to $\pi$, the region (II) in which the limit cycle is found tends to become smaller and smaller, and eventually completely disappears for $\phi=\pi$. Figure~\ref{fig:phasediagramdifferentphi} shows the two boundaries for the supercritical Hopf bifurcation from region (I) to region (II), and for the infinite-period bifurcation from region (II) to region (III) discussed in Fig.~\ref{fig:phasediagramwithflowphi0}, for different values of $\phi$. For $\phi=\pi$, the system directly passes from the below-threshold region to the synchronization one. Importantly, for the range of parameters considered here, the synchronization region after the limit cycle region, for small values of the coupling, is always found with four stable fixed points for all values of $\phi$.

Before concluding this section, we comment on the physicality of the model. As shown in Fig.~\ref{fig:phasediagramwithflowphi0}, our model predicts a large number of fixed points. Because we are describing a system of two parametric oscillators, one would expect that, in the synchronized regime, the spin picture holds when the system has only four stable fixed points on the real axis (twice as many as a single parametric oscillator). Indeed, for experimental purposes, the model that we consider is relevant only for values of the pump that are not too far away from the oscillations threshold (i.e., when the single oscillator has two stable fixed points, see Appendix~\ref{sec:decoupledparametricoscillators}). This is precisely the situation discussed in Ref.~\cite{PhysRevLett.123.083901}. In order for the model to be physical, it is important to verify that the condition $1-\beta({\bar x_1}^2+{\bar x_2}^2)/2>0$ holds inside the full phase diagram, where $\bar x^2_{1,2}$ is the long-time average of $x^2_{1,2}(t)$. This condition ensures that, on average, the energy of the pump is always down converted to the optical fields. We have verified that such condition holds inside the numerically explored phase diagram, even for very large values of the pump intensity.

Notice that the experimentally accessed region is usually up to $h/(2g)\simeq 2$, \rev{which corresponds to the maximum of the conversion efficiency (i.e., unity)}~\cite{Schiller:99,MARTINELLI2001,Sturman:11,Breunig2011,doi:10.1002/lpor.201600038}. To the best of our knowledge, the regions that are found for larger pump intensities, in which the model displays a number of stable points larger than four, remain experimentally unexplored. Our theory indicates the existence of interesting dynamics also in this high-pump intensity range. A deeper analysis of such regions is beyond the aim of the present manuscript and remains a subject of future studies.

\subsection{Critical scaling by a three-scale analysis}
\label{sec:criticalscalingbythreescaleanalysis}
In this section, we determine the critical exponent for the radius of the limit cycle close to the supercritical Hopf bifurcation boundary between regions (I) and (II), and of the period of the limit cycle close to the infinite-period bifurcation between region (II) and (III) of the phase diagram in Fig.~\ref{fig:phasediagramwithflowphi0}.

\begin{figure}[t!]
\centering
\includegraphics[width=8.2cm]{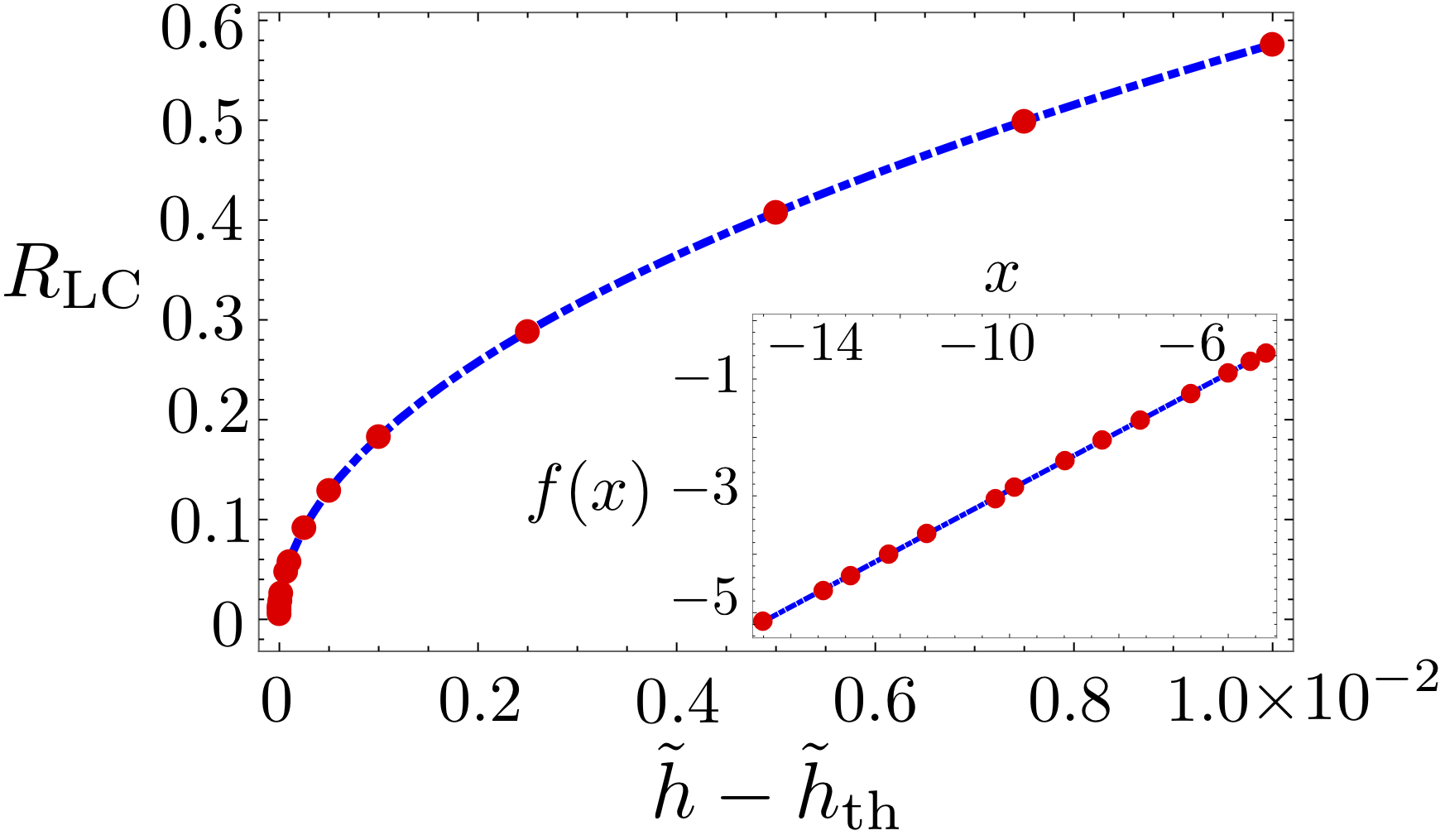}
\caption{Radius of the limit cycle $R_{\rm LC}$ as a function of $\tilde h-\tilde h_{\rm th}$, where $\tilde h_{\rm th}=2$. Red points are numerically obtained by solving Eq.~\eqref{eq:nonlinearmathieuequationflow2220coupledphihfixed} for $\phi=0$, $\tilde r=0.25$ and $\beta=10^{-2}$, and for $\tilde h$ close to the boundary of the supercritical Hopf bifurcation $\tilde h_{\rm th}=2$. We superimpose to the numerical data the analytic behaviour (blue dash-dotted line) found by the three-scale analysis in Eq.~\eqref{eq:radiuslimitcyclephi0case}. In order to further highlight the $1/2$ critical exponent of the supercritical Hopf bifurcation, we rescale the data by defining $x=\ln(1-\tilde h_{\rm th}/\tilde h)$ and, from Eq.~\eqref{eq:radiuslimitcyclephi0case}, $f(x)=\ln\sqrt{2/(3\beta)}+x/2$ (inset).}
\label{fig:radiuslimitcycle}
\end{figure}

In order to determine the scaling of the radius of the limit cycle, we focus on the points in region (II) which are close to the threshold $\tilde h_{\rm th}=2$ and sufficiently far from the infinite-period bifurcation. In this case, the condition $h\ll r\ll1$ holds and the dynamics of the oscillators is determined by three different characteristic frequencies: $\omega_0$ (fast oscillations), $\omega_0r$ (medium-scale beats) and $\omega_0h$ (long-time overall amplitude). In this condition, it is natural to perform a multiple-scale analysis by introducing three different time scales. If we redefine $r=r_0\,\tilde r$ and $h=g\,\tilde h$, where $r_0$ is a characteristic scale for the dynamics of the beats, we can distinguish three different time scales in the expansion: $t$ (for fast oscillations), $\sigma=r_0t$ (for the medium-scale dynamics, i.e., beats) and $\tau=gt$ (for the slow dynamics).

Focusing on the resonant case $\gamma=2\omega_0$, it is convenient to rewrite the equations of motion in Eq.~\eqref{eq:coupledparametricoscillatorequationswithnonlinearitydynamicalphigfixed} using the $x_\pm=x_1\pm i\,x_2$ basis, as in Sec.~\ref{sec:linearparametricoscillatorbyfloquetansatz}:
\begin{equation}
\begin{array}{l}
\ddot x_++\omega^2_0[1+g\tilde h\sin(2\omega_0t)]x_++\omega_0g\,\dot x_++i\,\omega_0r_0\tilde r\,\dot x_+\vspace{0.1cm}\\
\hspace{1cm}-\cfrac{\beta g\tilde h\omega_0^2}{4}\,\sin(2\omega_0t)\,\left(x^3_-+3\,x^2_+x_-\right)=0\vspace{0.1cm}\\
\ddot x_-+\omega^2_0[1+g\tilde h\sin(2\omega_0t)]x_-+\omega_0g\,\dot x_--i\,\omega_0r_0\tilde r\,\dot x_-\vspace{0.1cm}\\
\hspace{1cm}-\cfrac{\beta g\tilde h\omega_0^2}{4}\,\sin(2\omega_0t)\left(x^3_++3\,x_+x_-^2\right)=0
\end{array} \,\, .
\end{equation}
We now expand $x_\pm=x^{(0)}_\pm+r_0\,x^{(B)}_\pm+g\,x^{(1)}_\pm$. By proceeding as in Sec.~\ref{sec:coupledparametricoscillators} (see also Appendix~\ref{sec:detailsofhtederivation}), one obtains the equations for the fast-varying and medium-scale modes:
\begin{subequations}
\begin{align}
&\frac{\partial^2}{\partial t^2}\,x^{(0)}_\pm+\omega^2_0\,x^{(0)}_\pm=0 \label{eq:threescale1}\\
&\frac{\partial^2}{\partial t^2}x^{(B)}_\pm+2\,\frac{\partial}{\partial\sigma}\,\frac{\partial}{\partial t}\,x^{(0)}_\pm+\omega^2_0x^{(B)}_\pm\pm i\,\omega_0\tilde r\,\frac{\partial}{\partial t}x^{(0)}_\pm=0 \,\, ,\label{eq:threescale2}
\end{align}
\label{eq:threescale12}
\end{subequations}
and similarly the equations for the slow-varying modes
\begin{eqnarray}
&&\hspace{-0.4cm}\frac{\partial^2}{\partial t^2}\,x^{(1)}_\pm+2\,\frac{\partial}{\partial\tau}\,\frac{\partial}{\partial t}\,x^{(0)}_\pm+\omega^2_0\,x^{(1)}_\pm\nonumber\\
&&\hspace{-0.4cm}+\tilde h\omega^2_0\sin(2\omega_0t)\,x^{(0)}_\pm+\omega_0\frac{\partial}{\partial t}x^{(0)}_\pm\nonumber\\
&&\hspace{-0.4cm}-\frac{\beta\omega_0^2\tilde h}{4}\,\sin(2\omega_0t)\left[{\left(x_\mp^{(0)}\right)}^3+3{\left(x_\pm^{(0)}\right)}^2x^{(0)}_\mp\right]=0 \,\, .
\label{eq:threescale3}
\end{eqnarray}
From Eqs.~\eqref{eq:threescale12} and~\eqref{eq:threescale3}, since $x_\pm=x_\mp^*$ and since the equations for the slow-varying modes of $x_+$ and $x_-$ are mutually complex conjugated, we can use solutions of the form
\begin{equation}
\begin{array}{l}
x^{(0)}_+(t,\sigma,\tau)=C_+(\sigma)\left[A_S(\tau)e^{i\omega_0t}+A_S^*(\tau)e^{-i\omega_0t}\right]\vspace{0.1cm}\\
x^{(0)}_-(t,\sigma,\tau)=C_-(\sigma)\left[A_S(\tau)e^{i\omega_0t}+A_S^*(\tau)e^{-i\omega_0t}\right]
\end{array} \,\, ,
\label{eq:threescale4}
\end{equation}
where $C_\pm=C_\mp^*$ describe the medium-scale modes, and the slow-varying modes are described by the same complex amplitude $A_S$. By plugging these expressions into Eq.~\eqref{eq:threescale2}, one has the solvability condition for the medium-scale dynamics:
\begin{equation}
\frac{\partial}{\partial\sigma}\,C_\pm=\mp i\,\frac{\omega_0\tilde r}{2}\,C_\pm \,\, ,
\end{equation}
from which one obtains the beating factor $C_\pm(\sigma)=C(0)e^{\mp i\omega_0\tilde r\sigma/2}=C(0)e^{\mp i\omega_0rt/2}$, where $C(0)=|C(0)|e^{iu}$ is a complex number. Here, $|C(0)|$ is a normalization factor and $u$ determines the initial phase of the beats. By using these expression for $x_\pm^{(0)}$ in Eq.~\eqref{eq:threescale3}, and by neglecting oscillating factors as $e^{\pm i2\omega_0rt}$ that are strongly oscillating on the slow time scale, one obtains the solvability condition for the slow-varying amplitude $A_S$:
\begin{equation}
2\frac{\partial A_S}{\partial\tilde\tau}-\frac{\tilde h}{2}\,A_S^*+i\,A_S+\frac{3\tilde h\beta}{8}\left(3{|A_S|}^2A_S^*-A_S^3\right)=0 \,\, ,
\label{eq:coupledparametricoscillatorthreescaleanalysisd}
\end{equation}
which is nothing but Eq.~\eqref{eq:nonlinearmathieuequationflow2220coupledphihfixeda} without the $B$ term, and with the replacement $\beta\rightarrow3\beta/4$. For its solution, the reader is referred to Appendix~\ref{sec:decoupledparametricoscillators}. For completeness, we recall below the main results.

Basing on the notation used in Appendix~\ref{sec:decoupledparametricoscillators}, close to the threshold value $\tilde h\simeq2$, the origin is a saddle point and other four fixed points (two saddle points and two stable nodes) are found in its surroundings. Such points appear on the imaginary and real axes, respectively. We call the stable fixed point $W'_+$, whose polar coordinates in the ${\rm Im}(A_S)$ vs. ${\rm Re}(A_S)$ plane are [Eq.~\eqref{eq:radialcoordinatesingleoscillator}] $\varphi_{W'_+}=0$ and $R_{W'_+}=\sqrt{(2-4/\tilde h)/(3\beta)}$. Such a point is the only one that can stabilize the long-time dynamics of $x_\pm$. In particular, its coordinates determine the amplitude of the beats. Therefore, by recalling that, for $\phi=0$, $\tilde h_{\rm th}=2$, the radius of the limit cycle from the long-time dynamics of $A_S$, which we call $R_{\rm LC}\coloneqq\lim_{t\rightarrow\infty}|A_S(gt)|$, is readily determined:
\begin{equation}
R_{\rm LC}=\sqrt{\frac{2}{3\beta}\left(1-\frac{\tilde h_{\rm th}}{\tilde h}\right)} \,\, .
\label{eq:radiuslimitcyclephi0case}
\end{equation}
At the onset of the supercritical Hopf bifurcation, the limit cycle grows from zero amplitude with the critical exponent $1/2$. In terms of the $A$ and $B$ amplitudes determined by Eq.~\eqref{eq:nonlinearmathieuequationflow2220coupledphihfixed}, the limit cycle is therefore identified by the dynamics
\begin{subequations}
\begin{align}
\hspace{-0.2cm}|A(t)|&=|{\rm Re}[C_+(\sigma)]\,A_S(\tau)|=\left|\cos\left(\frac{\omega_0rt}{2}+u\right)\right|\,R_{LC}\\
\hspace{-0.2cm}|B(t)|&=|{\rm Im}[C_+(\sigma)]\,A_S(\tau)|=\left|\sin\left(\frac{\omega_0rt}{2}+u\right)\right|\,R_{LC} \,\, .
\end{align}
\label{eq:aandbinthelongtimelimit}
\end{subequations}
In the long-time limit, the limit cycle is therefore a perfect circular arc whose frequency is determined by $r$ and whose radius solely depends on $(\tilde h-\tilde h_{\rm th})/\tilde h$ [Eq.~\eqref{eq:radiuslimitcyclephi0case}].

In order to explicitly show the critical exponent $1/2$ of the supercritical Hopf bifurcation, we compute the radius of the limit cycle as a function of $\tilde h$ for a fixed $\tilde r\neq0$ by numerically solving Eq.~\eqref{eq:nonlinearmathieuequationflow2220coupledphihfixed} for $\phi=0$ and $\beta=10^{-2}$ close to the boundary of the supercritical Hopf bifurcation $\tilde h_{\rm th}=2$ (red dash-dotted line in Fig.~\ref{fig:phasediagramwithflowphi0}). The result is shown in Fig.~\ref{fig:radiuslimitcycle}. We then superimpose the numerically determined data with the expected behaviour [Eq.~\eqref{eq:radiuslimitcyclephi0case}] found by the three-scale analysis. The agreement between the two behaviours confirms the prediction found in Eq.~\eqref{eq:radiuslimitcyclephi0case}. Such square-root scaling can be further highlighted by rescaling the data and the analytical behaviour by introducing $x=\ln(1-\tilde h_{\rm th}/\tilde h)$ and, from Eq.~\eqref{eq:radiuslimitcyclephi0case}, $f(x)=\ln\sqrt{2/(3\beta)}+x/2$, which is shown in the inset.

\begin{figure}[t!]
\centering
\includegraphics[width=7.9cm]{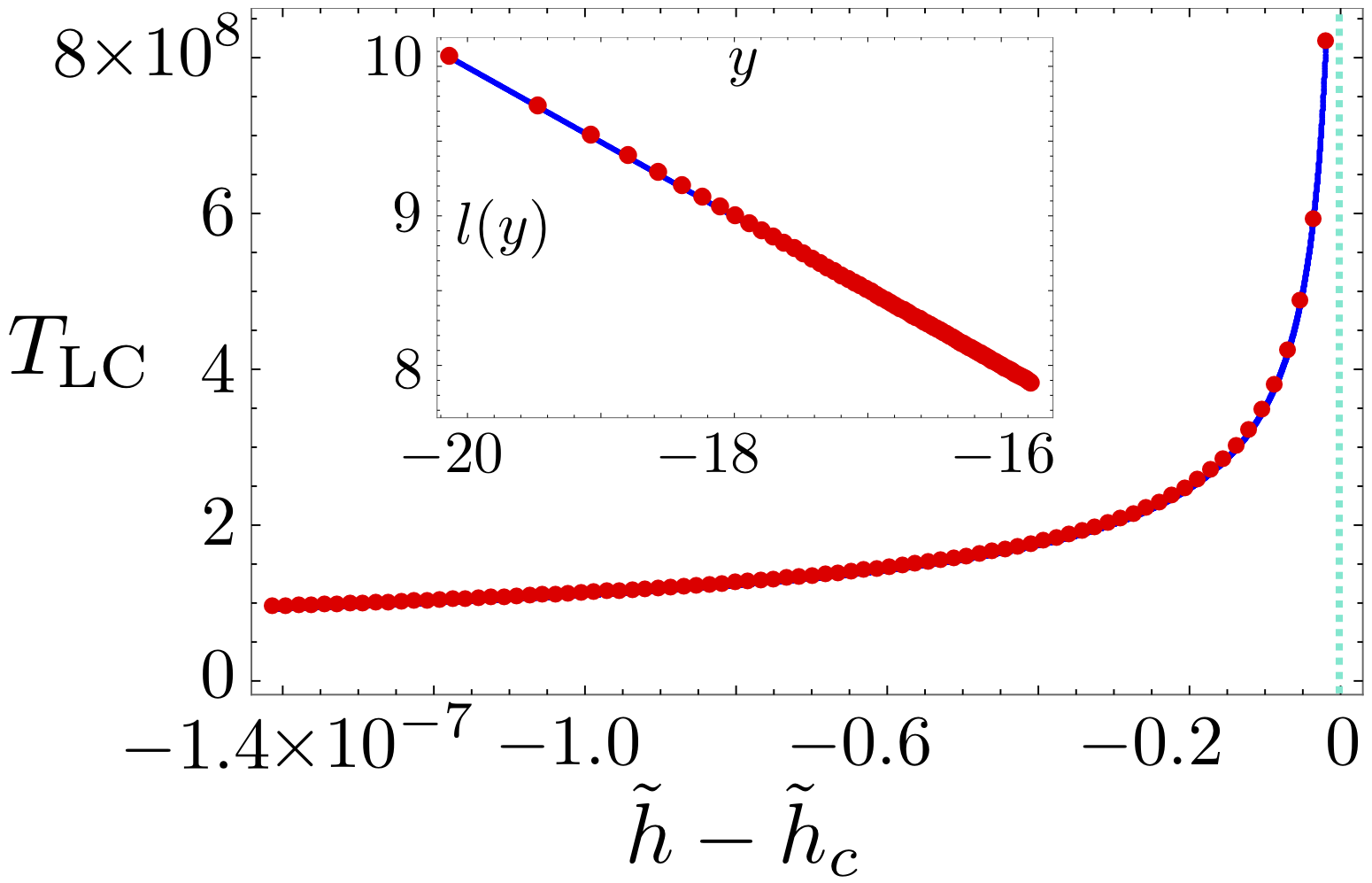}
\caption{Period of the limit cycle $T_{\rm LC}$ as a function of $\tilde h-\tilde h_c$, where $\tilde h_c$ is the critical value for the infinite-period bifurcation, which is numerically found: $\tilde h_{c}=2.02828427$. Red points are numerically obtained by solving Eq.~\eqref{eq:nonlinearmathieuequationflow2220coupledphihfixed} for $\phi=0$, $\beta=10^{-2}$, $\tilde r=5\times10^{-3}$ and by finely scanning $\tilde h$ close to the boundary of the infinite-period bifurcation (blue dashed line of the phase diagram in Fig.~\ref{fig:phasediagramwithflowphi0}). The data are fitted with the function $T_{\rm LC,fit}=c_1+c_2/\sqrt{\tilde h_c-\tilde h}$ (blue line), where $c_1$ and $c_2$ are fit parameters that are determined numerically. (Inset) Rescaled data by defining $y=\ln(\tilde h_{c}-\tilde h)$ and $l(y)=\ln(T_{\rm LC}-c_1)-\ln(c_2)=-y/2$ in order to highlight the exponent $1/2$. The cyan dashed vertical line highlights the phase boundary $\tilde h-\tilde h_c=0$.}
\label{fig:periodlimitcycle}
\end{figure}

The same critical exponent is found by studying the behaviour of the period of the limit cycle $T_{\rm LC}$ close to the infinite-period bifurcation~\cite{strogatz2007nonlinear} (blue dash line in the phase diagram in Fig.~\ref{fig:phasediagramwithflowphi0}). The result of the simulation is shown in Fig.~\ref{fig:periodlimitcycle}. We show the numerically determined value of $T_{\rm LC}$ as a function of the distance from the critical line $\tilde h-\tilde h_c$ at a fixed $\tilde r=5\times10^{-3}$. The continuous line represents the best fit of the form $T_{\rm LC,fit}=c_1+c_2/\sqrt{\tilde h_c-\tilde h}$, where $c_1$ and $c_2$ are fit parameters. As evident from the figure, the agreement between the numerical data and the fit confirms the fact that the period of the limit cycle diverges as $T_{\rm LC}\sim{(\tilde h_c-\tilde h)}^{-1/2}$ as the infinite-period bifurcation is approached.

Before concluding this section, we mention that the whole analysis remains valid if a different form of coupling is considered, i.e., $\omega_0^2r\,x_{2,1}$ in the equation of motion~\eqref{eq:coupledparametricoscillatorequationswithnonlinearitydynamicalphigfixed} of $x_1$ and $x_2$, respectively~\cite{PhysRevA.88.063853}. The proof of this statement is discussed in Appendix~\ref{sec:inphasecouplingmultiplescaleanalysis}.

\begin{figure}[t]
\centering
\includegraphics[width=8.5cm]{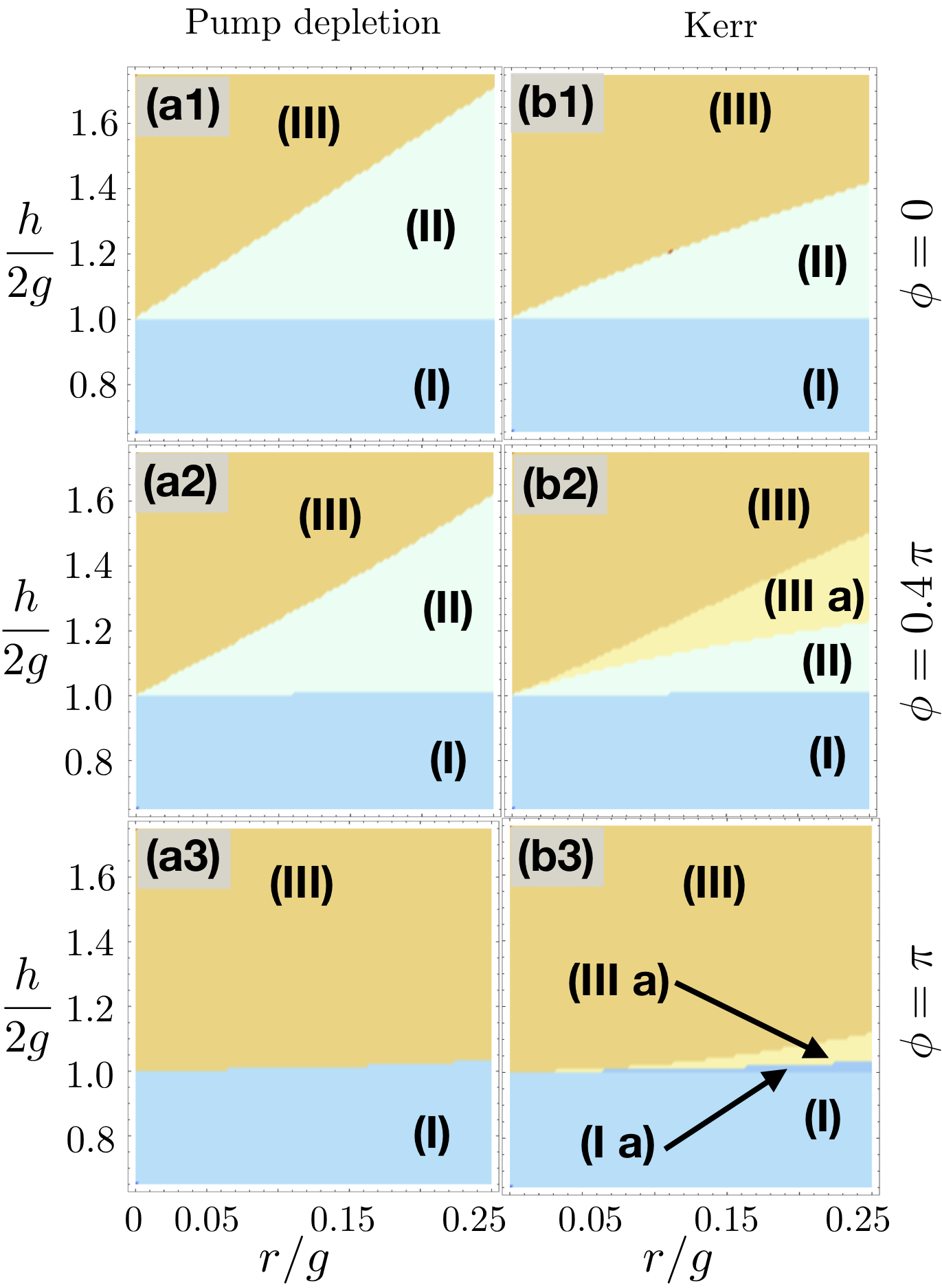}
\caption{Phase diagrams as in Fig.~\ref{fig:phasediagramwithflowphi0} for [(a1) and (b1)] $\phi=0$, [(a2) and (b2)] $\phi=0.4\,\pi$, and [(a3) and (b3)] $\phi=\pi$. The phase diagrams have been obtained by considering [panels (a)] pump-depletion nonlinearity, and [panels (b)] for Kerr nonlinearity. The different phases are: (I) phase in which the origin is the only stable attractor; (I a) phase in which the origin is a stable attractor and coexists with other attractors, no limit cycle is found in this phase; (II) phase with stable limit cycle and no stable attractor; (III) phase with four stable attractors, in which the origin is a saddle point; (III a) like with phase (III) but with two stable attractors only.}
\label{fig:phasediagramwithdifferentmodels}
\end{figure}

\section{Different types of nonlinearity}
\label{sec:discussionsondifferentmodels}
In this section, we comment on the effects of different nonlinearity of the two oscillators. We show that the phenomenology discussed in Sec.~\ref{sec:nonlinearcasebymultiplescaleanalysis} is not a consequence of the specific choice of the model, but it is common to other models which can be relevant in different experimental contexts. The properties that we discuss in this section are found by using exactly the same tools discussed in the previous sections. We therefore report the main results without explicitly showing all the technical details.

In order to ease the notation, we rewrite the equations of motion in a more compact and generic form as
\begin{equation}
\begin{array}{l}
\vspace{-0.2cm}
\ddot x_1+\omega^2_0\left[1+h\,\sin(\gamma t)\right]x_1+\omega_0g\,\dot x_1\\\\
\hspace{3.5cm}-\omega_0\,r\,\dot x_2+F_{\rm NL;1}=0\\\\
\vspace{-0.2cm}
\ddot x_2+\omega^2_0\left[1+h\,\sin(\gamma t+\phi)\right]x_2+\omega_0g\,\dot x_2\\\\
\hspace{3.5cm}+\omega_0\,r\,\dot x_1+F_{\rm NL;2}=0
\end{array} \,\, ,
\label{eq:coupledparametricoscillatorequationsgeneric}
\end{equation}
in which $F_{\rm NL;1,2}$ identify the nonlinear terms. In Eq.~\eqref{eq:coupledparametricoscillatorequationswithnonlinearitydynamicalphigfixed}, we considered the pump-depletion nonlinearity $F_{\rm NL,1}=-h\beta\omega^2_0\sin(2\omega_0t)\,x^3_1$, $F_{\rm NL,2}=-h\beta\omega^2_0\sin(2\omega_0t+\phi)\,x^3_2$. As mentioned before, another possible nonlinearity arises from a Kerr or saturation effect. In this case, the cubic term in Eq.~\eqref{eq:coupledparametricoscillatorequationswithnonlinearitydynamicalphigfixed} will not be coupled to the pump, and the equations of motion in Eq.~\eqref{eq:coupledparametricoscillatorequationsgeneric} are now written with $F_{\rm NL;1,2}=-\beta\omega^2_0\,x^3_{1,2}$ (see Table~\ref{tab:nonlinearitiesandcouplingsforcomparison}), whose multiple-scale equations are obtained as done for Eq.~\eqref{eq:nonlinearmathieuequationflow2220coupledphihfixed}.

The phase diagram that we obtain from the configurations of the fixed points, which is shown in Fig.~\ref{fig:phasediagramwithdifferentmodels}, panels (b), displays the same phases found for the phase diagram in Fig.~\ref{fig:phasediagramwithflowphi0}, which is reported for completeness in panels (a): (I) a region in which the system is below threshold, (II) an extended region in which only a stable limit cycle can stabilize the dynamics, therefore yielding also in this case everlasting beats in the time evolution of $x_1$ and $x_2$, and (III a)-(III) a region in which stable attractors stabilize the dynamics. However, differently from the model analyzed in Fig.~\ref{fig:phasediagramwithflowphi0}, in addition to region (III) in which four attractors are found, for $\phi>0$ [Fig.~\ref{fig:phasediagramwithdifferentmodels}, panel (b2)] there is an additional intermediate region, which we call region (III a), in which only \emph{two} attractors are found.

As discussed in Fig.~\ref{fig:phasediagramdifferentphi}, for $\phi=\pi$ [panels (a3) and (b3)], the region with the limit cycle [region (II)] disappears, and one passes directly from the below-threshold region to the region with stable attractors, which are four in the case of the pump-depletion nonlinearity and two in the case of the Kerr nonlinearity. In the latter case, for larger values of the pump, the region with four stable attractors is found above the one with two attractors only. In a more physical situation in which both nonlinearities are found, one always finds, for $\phi>0$, a small region with two stable fixed points before the one with four stable points, as the pump intensity is increased. Interestingly, when the Kerr nonlinearity is considered, the behaviour of the the radius of the limit cycle at the onset of the supercritical Hopf bifurcation (for $\phi=0$) is found to grow from zero with a critical exponent equal to $1/4$ in contrast to the critical exponent $1/2$ found in the case of the pump-depletion nonlinearity [see also Eq.~\eqref{eq:fixedpointkerr}]. Such difference can be exploited in experiments to distinguish between the two types nonlinearities. This point is left for future work.

\begin{table}[t]
\centering
\begin{tabular}{c}
\hline
\textbf{Nonlinearity} \\\hline
Pump depletion\\
$F_{\rm NL,1}=-h\beta\omega^2_0\sin(2\omega_0t)\,x^3_1$\\
$F_{\rm NL,2}=-h\beta\omega^2_0\sin(2\omega_0t+\phi)\,x^3_2$\\\hline
Kerr, saturation\\
$F_{\rm NL;1}=-\beta\omega^2_0\,x^3_{1}$\\
$F_{\rm NL;2}=-\beta\omega^2_0\,x^3_{2}$\\\hline
\end{tabular}
\caption{Types of nonlinearity (pump depletion or Kerr/saturation) that we consider in the equations of motion in Eq.~\eqref{eq:coupledparametricoscillatorequationsgeneric}.}
\label{tab:nonlinearitiesandcouplingsforcomparison}
\end{table}

From this analysis, apart from the specific quantitative details that depend on the specific model that we consider, it is therefore seen that the presence of a wide region in which the system displays everlasting beats comes solely from the interplay between parametric gain, losses, nonlinearity, coupling and, apart from extremely fine-tuned phase differences between the two oscillators, it always emerges for any nonzero coupling as the oscillation threshold is crossed (see Sec.~\ref{sec:dissipativecouplngandcim} for the extension to the case of dissipative coupling). When the system is within such phase, the system of two coupled parametric oscillator therefore does not match the picture of two Ising spins discussed in previous work~\cite{PhysRevA.88.063853,nphoton.2016.68,s41534-017-0048-9,1805.05217}.

\section{Dissipative coupling and CIM}
\label{sec:dissipativecouplngandcim}
\rev{So far, we considered only the effect of an energy-preserving coupling, which is found when the energy exchange rate between the two oscillator is balanced.} In this section, we consider the effects of a dissipative coupling, \rev{which is instead found whenever the two oscillators exchange energy with different rates}, in order to connect our model to the one discussed in Ref.~\cite{PhysRevA.88.063853} in the context of CIMs. We eventually a generic experimental implementation of such a coupling.

In order to show this connection, we first consider the linear case. We rewrite Eq.~\eqref{eq:coupledparametricoscillatorequationslinearphi2} as
\begin{equation}
\begin{array}{l}
\ddot x_1+\omega^2_0\left[1+h\,\sin(\gamma t-\phi/2)\right]x_1+\omega_0g\,\dot x_1\vspace{0.1cm}\\
\hspace{2cm}-\omega_0(r-\alpha)\,\dot x_2=0\vspace{0.1cm}\\
\ddot x_2+\omega^2_0\left[1+h\,\sin(\gamma t+\phi/2)\right]x_2+\omega_0g\,\dot x_2\vspace{0.1cm}\\
\hspace{2cm}+\omega_0(r+\alpha)\,\dot x_1=0
\end{array} \,\, .
\label{eq:coupledparametricoscillatorequationslinearphidissipativecoupling}
\end{equation}
where all quantities are as in Eq.~\eqref{eq:coupledparametricoscillatorequationslinearphidissipativecoupling}, and $\alpha\geq0$ represents the strength of the dissipative part of the coupling \rev{that quantifies the unbalancing between the energy exchange rates between the two oscillators}. As we will show, depending on the relation between $r$ and $\alpha$, the system undergoes a transition between the CIM behaviour~\cite{PhysRevA.88.063853} and the beating phenomenology discussed in the previous sections.

In order to diagonalize Eq.~\eqref{eq:coupledparametricoscillatorequationslinearphidissipativecoupling}, we introduce the basis ${(k_-\,\,k_+)}^T=K(x_1\,\,x_2)^T$, where $T$ denotes the transposition and the non-unitary matrix $K$ is
\begin{equation}
K\!=\!\frac{1}{\sqrt{1+\left|(r-\alpha)/(r+\alpha)\right|}}\left(\begin{array}{cc}1&-i\sqrt{(r-\alpha)/(r+\alpha)}\\1&i\sqrt{(r-\alpha)/(r+\alpha)}\end{array}\right)
\end{equation} 
In this basis, the equation of motion~\eqref{eq:coupledparametricoscillatorequationslinearphidissipativecoupling} becomes
\begin{eqnarray}
&&\hspace{-0.7cm}\ddot k_\pm+\omega^2_0\left[1+h\,\sin(\gamma t)\cos(\phi/2)\right]k_\pm+\omega_0\,g\,\dot k_\pm\nonumber\\
&&\hspace{-0.7cm}\mp i\,\omega_0\sqrt{r^2-\alpha^2}\,\dot k_\pm-h\omega_0^2\sin(\phi/2)\,\cos(\gamma t)\,k_\mp=0 \,\, .
\label{eq:coupledparametricoscillatorequationslinearphidissipativecoupling2}
\end{eqnarray}
From Eq.~\eqref{eq:coupledparametricoscillatorequationslinearphidissipativecoupling2} and using the same tools as in Sec.~\ref{sec:linearparametricoscillatorbyfloquetansatz}, one can see that two different regimes arise:
\begin{enumerate}
\item For $r>\alpha$, i.e., when the nature of the coupling is mostly non dissipative, the term $\sqrt{r^2-\alpha^2}$ is real. This is the situation studied in Sec.~\ref{sec:linearparametricoscillatorbyfloquetansatz}. When $\gamma=2\omega_0$, the solution displays beats at a frequency $\pm\omega_0\sqrt{r^2-\alpha^2}/2$. Two additional parametric resonances at $\gamma=2\omega_0\pm\omega_0\sqrt{r^2-\alpha^2}$ are found, at which parametric amplification occurs without beats;
\item For $r<\alpha$, i.e., when the dissipative part of the coupling dominates, the term $\sqrt{r^2-\alpha^2}=i\,\sqrt{\alpha^2-r^2}$ is imaginary. Now, only the parametric resonance at $\gamma=2\omega_0$ is found, for all values of $\phi$, and the solution never displays beats. Instead, the modes $k_\pm$ have now different loss terms $g\mp\sqrt{\alpha^2-r^2}$ for $k_\pm$, leading to different oscillation thresholds.
\end{enumerate}
\rev{The systems therefore undergoes a transition between the CIM to the beating behaviour at $r=\alpha$}. The analysis can be extended to the non-linear case by including the pump-depletion nonlinearity, and proceeding with the two-scale expansion as in Eq.~\eqref{eq:nonlinearmathieuequationflow2220coupledphihfixed}, see Appendix~\ref{sec:oscilaltionthresholdandbeats} for more details. By using this method, we compute the phase diagram in the $h/(2g)$ vs. $r/g$ plane, as done in Figs.~\ref{fig:phasediagramwithflowphi0} and~\ref{fig:phasediagramwithdifferentmodels}, for different values of $\alpha$ (which we also rescale as $\tilde\alpha=\alpha/g$) and $\phi$.
\begin{figure}[t]
\centering
\includegraphics[width=8.5cm]{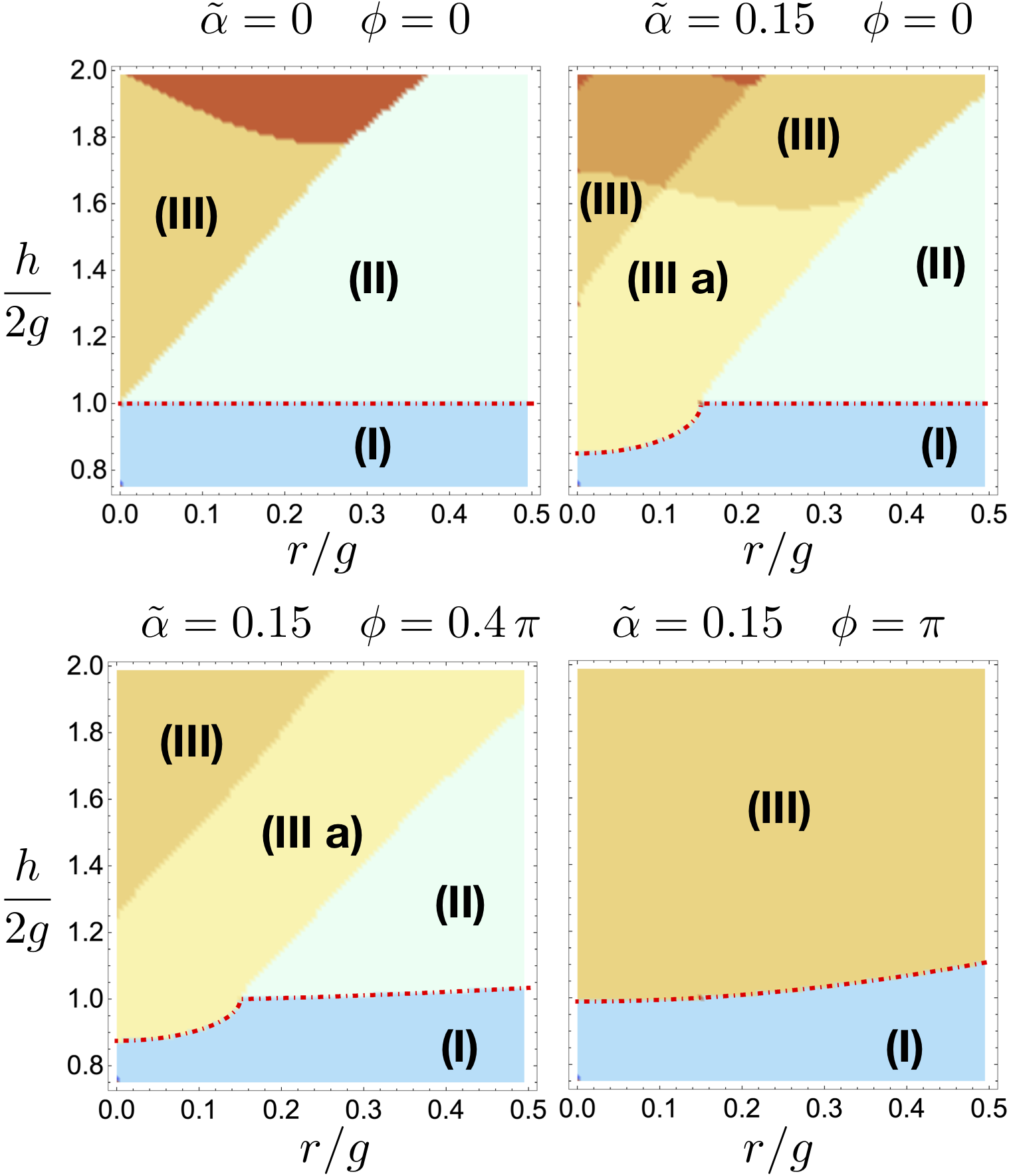}
\caption{Phase diagram in the $h/(2g)$ vs. $r/g$ plane as in Figs.~\ref{fig:phasediagramwithflowphi0} and~\ref{fig:phasediagramwithdifferentmodels}, for different values of $\tilde\alpha$ and $\phi$, as in the legends. For $\tilde\alpha>0$, in addition to (I) below threshold region, (II) limit cycle region, (III) region with four stable fixed points and phases with more than four fixed points (not labelled), which are not relevant for the present purpose, a region (III a) with two stable fixed points arises: for $\tilde r<\tilde\alpha$ right above threshold, and for $\tilde r>\tilde\alpha$, after the limit cycle region. The former case is the working point for the two-oscillator CIM, see also Ref.~\cite{PhysRevA.88.063853}. The red dashed line is the analytical threshold (Appendix~\ref{sec:oscilaltionthresholdandbeats}).}
\label{fig:phasediagramalphaphi}
\end{figure}
The result is shown in Fig.~\ref{fig:phasediagramalphaphi}. The phase diagram for $\tilde\alpha=0$ and $\phi=0$ is the same as in Figs.~\ref{fig:phasediagramwithflowphi0} and~\ref{fig:phasediagramwithdifferentmodels}, panel (a1), and is reported here for completeness. As explained there, when crossing the \rev{oscillation} threshold $h=2g$, one always enters the beating region. For stronger pumps, the system undergoes a transition to a region with four (or eight) fixed points. The picture changes when $\tilde\alpha>0$. In particular, we choose $\tilde\alpha=0.15$ and first show the result for $\phi=0$. We see that, in this case, an additional phase with \emph{two} stable fixed points emerges, and for $\tilde r<\tilde\alpha$, this phase is found directly above the threshold (see Appendix~\ref{sec:oscilaltionthresholdandbeats} for the analytical computation). For larger values of $\tilde h$, from the region with two stable points, the phase with four stable fixed points is found. This situation matches the one discussed in Ref.~\cite{PhysRevA.88.063853}, in which the two-oscillator system can be used as a CIM directly above the oscillation threshold. For $\tilde r>\tilde\alpha$, the limit cycle region discussed throughout this manuscript emerges between the below-threshold and the CIM regions. This analysis suggests that there are two different routes to reach the CIM regime, whose further analysis is left for future work.

For $\phi>0$, the picture remains qualitatively similar. The width of the region with two stable fixed points, as well as the limit cycle region, is reduced as $\phi$ is increased, as discussed in the previous sections for $\tilde\alpha=0$. For $\phi=\pi$, only the region with four stable fixed points is found above threshold.

\begin{figure}[t]
\centering
\includegraphics[width=8cm]{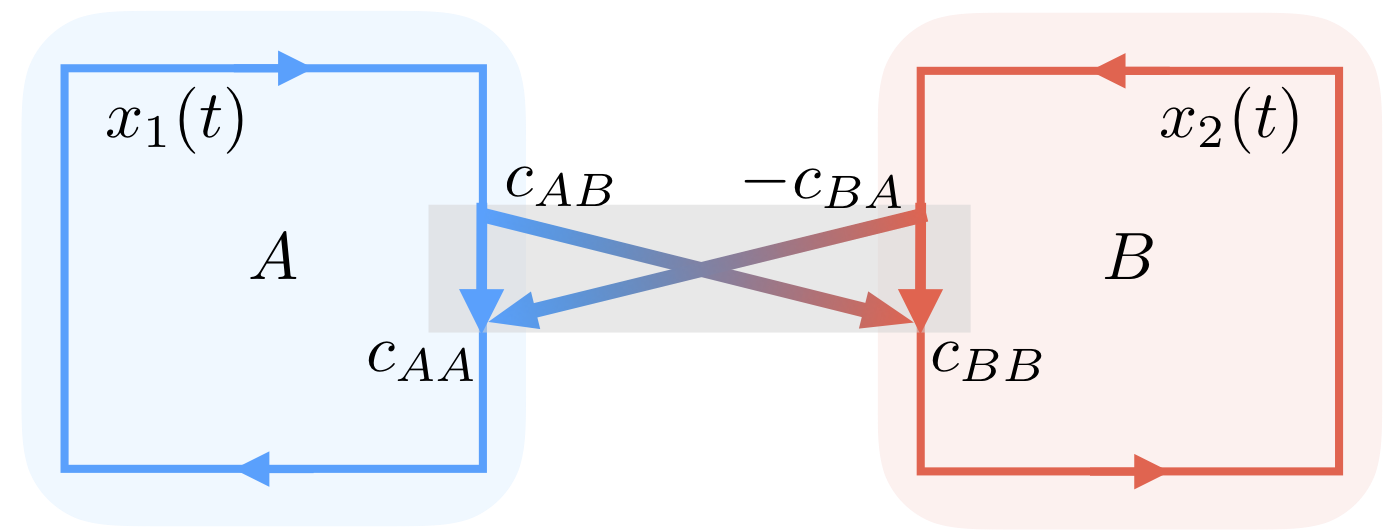}
\caption{\rev{Generic scheme for an experimental setup with power-splitter coupler. The transmittance and energy transfer coefficients are identified by $c_{AA},c_{BB},c_{AB},c_{BA}$. When the exchange channels identified by $c_{AB}$ and $c_{BA}$ are lossy causing \emph{different} coupling rates, i.e., $c_{AB}\neq c_{BA}$, the coupling as in Eq.~\eqref{eq:coupledparametricoscillatorequationslinearphidissipativecoupling} can be obtained.}}
\label{fig:figurescheme}
\end{figure}

\rev{Before concluding, we discuss a generic experimental setup to realize our system as in Eq.~\eqref{eq:coupledparametricoscillatorequationslinearphidissipativecoupling}. A minimal setup is reported in Fig.~\ref{fig:figurescheme}, which can be implemented both by means of radio-frequency~\cite{PhysRevLett.123.083901} or optical components, thus ensuring the scalability of the setup. The two fields $x_1$ and $x_2$ are generated inside two cavities $A$ and $B$, respectively, and they are coupled by a power-splitter coupling. In the most general case, this component accounts for the following quantities: (i) the transmittance coefficients $c_{AA}$ and $c_{BB}$, which can be without loss of generality taken equal for both cavities, i.e., $c_{AA}=c_{BB}=c\geq0$, whose effect is to renormalize the intrinsic loss of the cavities $g$, and (ii) the coupling coefficients $c_{AB}$ and $c_{BA}$ that when they have the same sign, determine the amount of energy that is transferred from $A$ to $B$, and from $B$ to $A$, respectively. Without loss of generality, we consider $c_{BA}>0$.}

\rev{In this notation, the coupling between the two oscillators can be written as
\begin{equation}
\begin{array}{l}\dot x_1=+\omega_0\,c_{AB}\,x_2\\\dot x_2=-\omega_0\,c_{BA}\,x_1\end{array} \,\, .
\label{eq:equationforthefieldsinsidethecavities211}
\end{equation}
When $c_{AB}=c_{BA}$, one defines $r=c_{AB}=c_{BA}$, and this balanced coupling leads only to the presence of beats, without any CIM region. Instead, when the coupling is unbalanced, i.e. $c_{AB}\neq c_{BA}$, it is possible to achieve the CIM regime. In this case, one can write $r=(c_{BA}+c_{AB})/2$ and $\alpha=(c_{BA}-c_{AB})/2$, so that $c_{AB}=r-\alpha$ and $c_{BA}=r+\alpha$, and Eq.~\eqref{eq:equationforthefieldsinsidethecavities211} becomes
\begin{equation}
\begin{array}{l}\dot x_1=+\omega_0\,(r-\alpha)\,x_2\\\dot x_2=-\omega_0\,(r+\alpha)\,x_1\end{array} \,\, .
\label{eq:equationforthefieldsinsidethecavities2112}
\end{equation}
By taking the time derivative on both sides of Eq.~\eqref{eq:equationforthefieldsinsidethecavities2112}, and by including this coupling in the equations of motion, Eq.~\eqref{eq:coupledparametricoscillatorequationslinearphidissipativecoupling} is obtained.} 

\vspace{-0.2cm}
\section{Conclusions}
\label{sec:conclusions}
In this work, we reported a detailed analytical and numerical analysis of two parametric oscillator coupled by a power-splitting coupling, first focusing on the case in which the coupling was purely energy-preserving, and later discussing the relation of our model with CIMs in the case of a dissipative coupling.

We first studied in detail the linear case by resorting to the Floquet theorem. We analytically showed that the system displays three resonances, whose relative splitting in frequency depends on the coupling strength, and then we numerically determined the full stability phase diagram. We showed that, depending on what resonance is met, parametric amplification for both oscillators can occur with or without the beats. In the former case, the frequency of the beats is solely determined by the coupling strength

We then discussed the nonlinear case, first by studying in detail the model with one specific type of nonlinearity, namely, the pump depletion. Next, we corroborated the generality of our finings by discussing the validity of our results in different models, considering different types of nonlinearity and coupling. A single parametric oscillator, above the oscillation threshold, has two possible solutions that are identified by a relative time shift of $\pi$ (one period of the pump). For this reason, a single parametric oscillation is suitable for the simulation of a classical spin-1/2 degrees of freedom, the two states of the spin being identified by the two solutions. In contrast, we showed that two nonlinear coupled parametric oscillators display a wide region in parameter space in which, sufficiently not too far away from the oscillation threshold, only a stable limit cycle is found and oscillations occurs with everlasting beats whose shape and frequency depends on the system parameters. This phenomenology was found as long as the nature of the coupling was mostly non-dissipative, irrespective of the details of the nonlinearity and away from extremely fine-tuned values of the phase difference between the pumps.

Our findings, \rev{from a generic perspective,} show a way to use parametric oscillators in order to preserve coherence indefinitely.
\rev{On the other hand, they are immediately relevant to the context of CIMs. Indeed, given the richer physics that we found in the minimal building block of the two-oscillator system with respect to what has been previously addressed, it is crucial to understand how the interplay between two couplings of different nature affects a more structured network.}
For this reason, the extension of the study presented in this manuscript to more than two coupled parametric oscillators, specifically, studying the fate of the limit cycle when several oscillations are coupled is an important step in the analysis of large-scale CIMs. We leave this point as an \rev{outstanding} perspective for future work.

\begin{center}
\begin{small}
\textbf{ACKNOWLEDGEMENTS}
\end{small}
\end{center}
We thank Joseph Avron, Ivan Bonamassa, Claudio Conti, Nir Davidson, Igor Gershenzon, Ron Lifshitz, Chene Tradonsky, and Yoshihisa Yamamoto for fruitful discussions. We are grateful to David A. Kessler for careful reading and invaluable comments on this manuscript. A.~P. acknowledges support from ISF grant No.~46/14.  M.~C.~S. acknowledges support from the ISF grants No.~231/14 and~1452/14.


\appendix

\section*{Appendix}
\renewcommand{\theequation}{A\arabic{equation}}
\setcounter{equation}{0}
\subsection{Details on the derivation of Eq.~\eqref{eq:nonlinearmathieuequationflow2220coupledphihfixed}}
\label{sec:detailsofhtederivation}
In this appendix, we report the details of the derivation of the system in Eq.~\eqref{eq:nonlinearmathieuequationflow2220coupledphihfixed}. As we discussed in the main text, one first separates the fast-varying time scale $t$ from the slow varying one $\tau=gt$. We now proceed with the perturbative expansion, treating $g$ as the small expansion parameter, and consider only terms in the expansion that are at most of the order of $g$. First, in the dynamics of $x_1$ and $x_2$, we can explicitly separate the fast-varying time scale from the slow-varying one, i.e., we write $x_{1,2}=x_{1,2}(t,\tau)$. We can therefore express the time derivative as $d/dt=\partial/\partial t+g(\partial/\partial\tau)$, and therefore $d^2/dt^2\simeq\partial^2/\partial\tau^2+2g(\partial/\partial\tau)(\partial/\partial t)$, where we neglect terms of the order of $g^2$. Similarly, we expand $x_{1,2}=x^{(0)}_{1,2}+g\,x^{(1)}_{1,2}$, where $x^{(0)}_{1,2}$ and $x^{(1)}_{1,2}$ represent the zero-order and first-order correction to $x_{1,2}$, respectively.

Using these definitions into Eq.~\eqref{eq:coupledparametricoscillatorequationswithnonlinearitydynamicalphigfixed} and the fact that $2\Omega_r\simeq2\omega_0$, we can separate the terms that do not appear multiplied by $g$, which are 
\begin{equation}
\frac{\partial^2}{\partial t^2}\,x^{(0)}_1+\omega^2_0\,x^{(0)}_1=0 \qquad \frac{\partial^2}{\partial t^2}\,x^{(0)}_2+\omega^2_0\,x^{(0)}_2=0 \,\, ,\label{eq:multiplescaleequationsystemphi1}
\end{equation}
from the terms that are proportional to $g$, which are
\begin{subequations}
\begin{align}
&\hspace{-0.2cm}\frac{\partial^2}{\partial t^2}\,x^{(1)}_1+\omega^2_0\,x^{(1)}_1+2\,\frac{\partial}{\partial\tau}\,\frac{\partial}{\partial t}\,x^{(0)}_1-\tilde r\,\frac{\partial}{\partial t}\,x^{(0)}_2+\omega_0\,\frac{\partial}{\partial t}\,x^{(0)}_1\nonumber\\
&\hspace{0.1cm}+\,\omega^2_0\,\tilde h\,\left[1-\beta{\left(x^{(0)}_1\right)}^2\right]\,\sin(2\omega_0t)\,x^{(0)}_1=0 \label{eq:multiplescaleequationsystemphi2}\\
&\hspace{-0.2cm}\frac{\partial^2}{\partial t^2}\,x^{(1)}_2+\omega^2_0\,x^{(1)}_2+2\,\frac{\partial}{\partial\tau}\,\frac{\partial}{\partial t}\,x^{(0)}_2+\tilde r\,\frac{\partial}{\partial t}\,x^{(0)}_1+\omega_0\,\frac{\partial}{\partial t}\,x^{(0)}_2\nonumber\\
&\hspace{0.1cm}+\,\omega^2_0\,\tilde h\,\left[1-\beta{\left(x^{(0)}_2\right)}^2\right]\,\sin(2\omega_0t+\phi)\,x^{(0)}_2=0  \,\, .\label{eq:multiplescaleequationsystemphi4}
\end{align}
\end{subequations}
From Eq.~\eqref{eq:multiplescaleequationsystemphi1}, we can write $x_1(t,\tau)=A(\tau)\,e^{i\omega_0 t}+A^*(\tau)\,e^{-i\omega_0t}$ and $x_2(t,\tau)=B(\tau)\,e^{i\omega_0 t}+B^*(\tau)\,e^{-i\omega_0t}$, where $A(\tau)$ and $B(\tau)$ represent the slow-varying complex amplitudes for $x_1$ and $x_2$, respectively. If these expressions of $x^{(0)}_{1,2}$ are used into Eq.~\eqref{eq:multiplescaleequationsystemphi2} and Eq.~\eqref{eq:multiplescaleequationsystemphi4}, one has [we show explicitly the calculation for Eq.~\eqref{eq:multiplescaleequationsystemphi2} only, the one for Eq.~\eqref{eq:multiplescaleequationsystemphi4} being essentially the same]
\begin{eqnarray}
&&\frac{\partial^2}{\partial t^2}\,x^{(1)}_1+\omega^2_0\,x^{(1)}_1+i\,e^{i\omega_0t}\left[2\omega_0\,\frac{\partial A}{\partial\tau}-\frac{\omega_0^2\tilde h}{2}\,A^*+\omega_0^2\,A\right.\nonumber\\
\nonumber\\
&&\hspace{0.2cm}\left.+\frac{\omega_0^2\tilde h\,\beta}{2}\left(3{|A|}^2A-A^3\right)-\omega_0^2\,\tilde r\,B\right]+{\rm c.c.}=0 \,\, ,
\label{eq:multiplescaleeuqation1}
\end{eqnarray}
where ${\rm c.c.}$ denotes the complex conjugation. The terms proportional to $e^{\pm i\omega_0t}$ in Eq.~\eqref{eq:multiplescaleeuqation1}, which are commonly referred to as secular terms, represent a resonant driving force applied to the $x^{(1)}_1$ oscillator. Such a force, will always cause the solution for $x^{(1)}_1$ to be unbounded. In order to ensure the solvability of Eq.~\eqref{eq:multiplescaleeuqation1}, we need to impose that such secular terms are zero. This gives the solvability condition for Eq.~\eqref{eq:multiplescaleeuqation1}:
\begin{eqnarray}
&&2\omega_0\,\frac{\partial A}{\partial\tau}-\frac{\omega_0^2\tilde h}{2}\,A^*+\frac{\omega_0^2\tilde h\,\beta}{2}\left(3{|A|}^2A^*-A^3\right)\nonumber\\
&&\hspace{3.5cm}+\omega_0^2\,A-\omega_0^2\,\tilde r\,B=0 \,\, .\\\nonumber
\label{eq:multiplescaleeuqation2a}
\end{eqnarray}
By separating real and imaginary part of $A$ and $B$, i.e., $A=A_R+i\,A_I$ and $B=B_R+i\,B_I$, we can write the two coupled equations for $A_R$ and $A_I$ shown in Eq.~\eqref{eq:nonlinearmathieuequationflow2220coupledphihfixeda}. By repeating the same steps for Eq.~\eqref{eq:multiplescaleequationsystemphi4}, we therefore arrive to the set of four coupled equations for the real and imaginary parts of the complex amplitudes of the fields in Eq.~\eqref{eq:nonlinearmathieuequationflow2220coupledphihfixed}.

\begin{figure}[t]
\centering
\includegraphics[width=8.5cm]{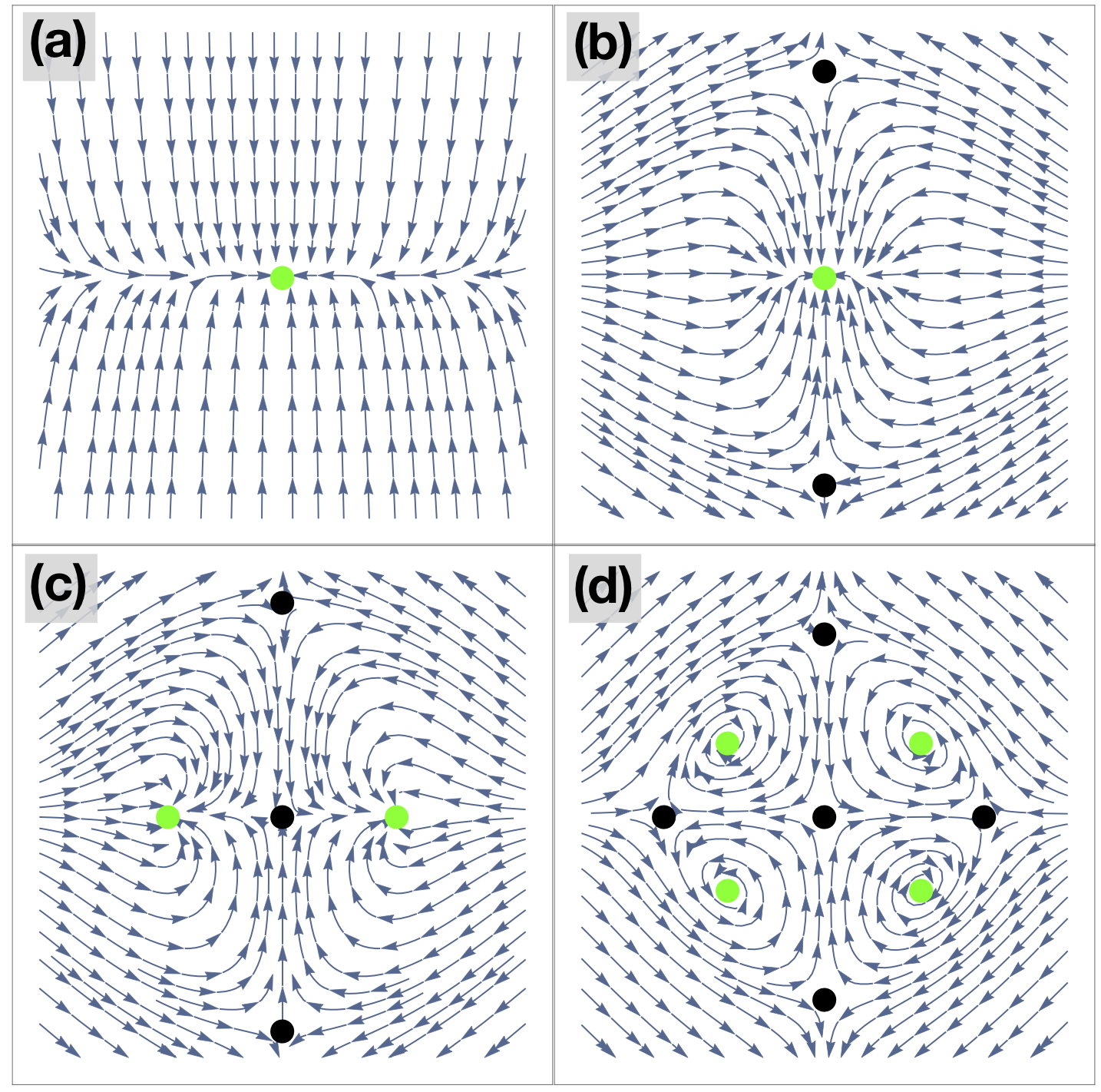}
\caption{Flow of the nonlinear Mathieu's equation with pump-depletion nonlinearity [Eq.~\eqref{eq:mathieruequationsnonlinearpumpdepletion}] in the $A_I$ vs. $A_R$ plane. Blue arrows represent the lines of the flow, black and green dots represent unstable and stable fixed point, respectively. We show the flow for four prototype cases: (a) for $\beta=0$ and $\tilde h<2$, (b) $\beta>0$ and $\tilde h<2$, (c) $\beta>0$ and $2<\tilde h<4$, and (d) $\beta>0$ and $\tilde h>4$.}
\label{fig:flowequationaphi0}
\end{figure}

\begin{figure}[t]
\centering
\includegraphics[width=8.5cm]{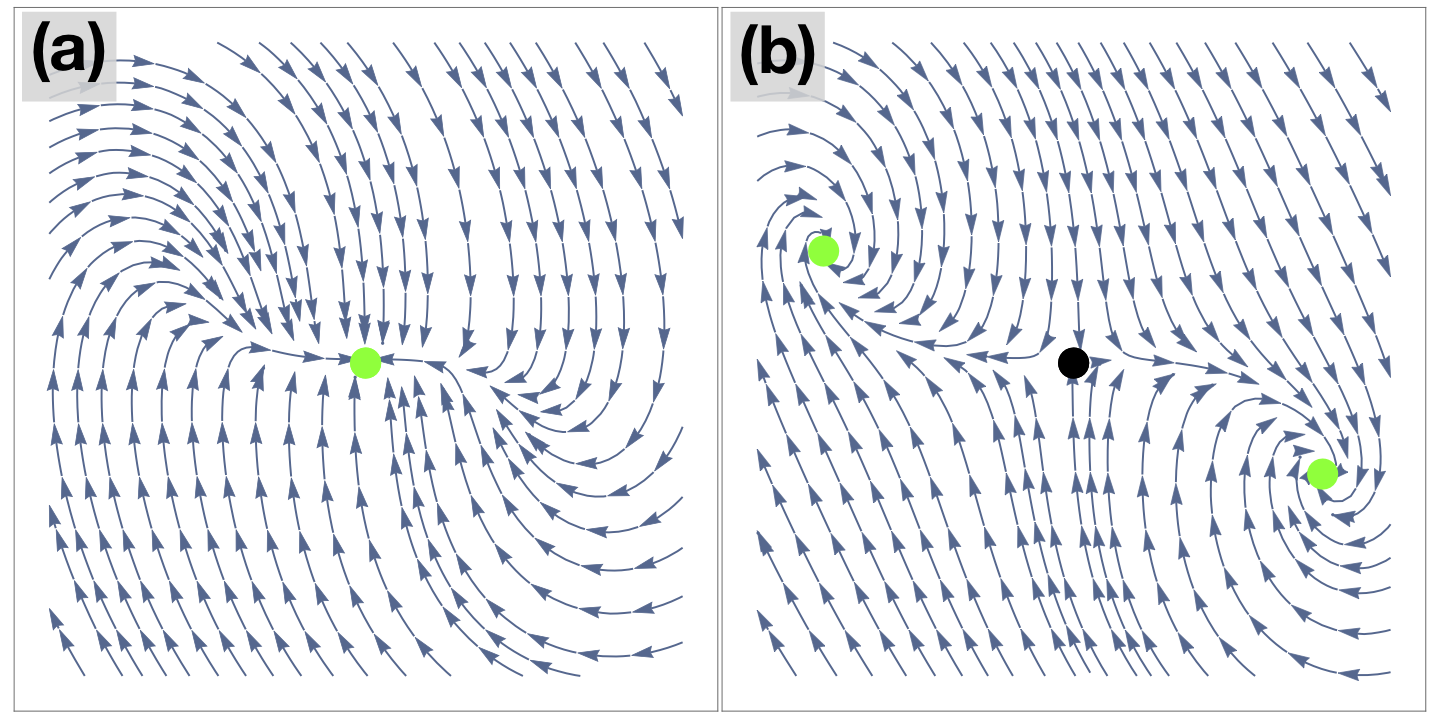}
\caption{Flow of the nonlinear Mathieu's equation with Kerr nonlinearity [Eq.~\eqref{eq:nonlinearmathieuequationflow22}] as in Fig.~\ref{fig:flowequationaphi0}. We show the flow for two prototype cases for $\beta>0$: (a) for $\tilde h<2$ and (b) $\tilde h>2$}
\label{fig:flowequationaphi1}
\end{figure}

\subsection{Stability analysis of the single parametric oscillator}
In this appendix, we report for completeness the stability analysis of the nonlinear Mathieu's equation for the single parametric oscillator in the presence of the pump-depletion or Kerr nonlinearity.

\subsubsection{Pump-depletion nonlinearity}
\label{sec:decoupledparametricoscillators}
We first focus on the case of the pump-depletion nonlinearity, i.e., the case discussed in Sec.~\ref{sec:coupledparametricoscillators}, for $\tilde r=0$. In this case, the two oscillators are decoupled and one can study only the dynamics of one of the two oscillators $A$ or $B$ in Eq.~\eqref{eq:nonlinearmathieuequationflow2220coupledphihfixed}, since for $r=0$ the effect of $\phi$ is trivial and the two equations of motion describe exactly the same physics. In order to simplify the analytical calculation, we therefore study the equations for $A$ in Eq.~\eqref{eq:nonlinearmathieuequationflow2220coupledphihfixed}, for $\phi=0$, which for completeness we recall below ($\tilde\tau=\omega_0\tau$):
\begin{subequations}
\begin{align}
&\frac{\partial A_R}{\partial\tilde\tau}=A_R\left[\frac{\tilde h}{4}-\frac{1}{2}-\frac{\beta\tilde h}{2}\,\left(A^2_R+3\,A^2_I\right)\right]\\
&\frac{\partial A_I}{\partial\tilde\tau}=A_I\left[-\frac{\tilde h}{4}-\frac{1}{2}+\frac{\beta\tilde h}{2}\,\left(A^2_I+3\,A^2_R\right)\right] \,\, .
\end{align}
\label{eq:mathieruequationsnonlinearpumpdepletion}
\end{subequations}
It is convenient to find the coordinates of the fixed points in the $A_I$ vs. $A_R$ plane in polar coordinates. We can therefore define $A_R=R\,\cos(\varphi)$ and $A_I=R\,\sin(\varphi)$. From Eq.~\eqref{eq:mathieruequationsnonlinearpumpdepletion}, the condition $\partial A/\partial\tilde\tau=0$ therefore yields the set of equations for $R\neq0$:
\begin{subequations}
\begin{align}
&\cos(\varphi)\left[\frac{\tilde h}{2}-1-\beta\tilde hR^2\left[\cos^2(\varphi)+3\sin^2(\varphi)\right]\right]=0  \label{eq:multiplescalepolarcoordinatesdynamic1p}\\
&\sin(\varphi)\left[-\frac{\tilde h}{2}-1+\beta\tilde hR^2\left[\sin^2(\varphi)+3\cos^2(\varphi)\right]\right]=0 \,\, .\label{eq:multiplescalepolarcoordinatesdynamic2p}
\end{align}
\label{eq:equationdforfixedpointsdynamicp}
\end{subequations}
From Eqs.~\eqref{eq:equationdforfixedpointsdynamicp}, one sees that $\varphi=0$ and $\varphi=\pi/2$ are two possible solutions for the angular variable, which define two sets of fixed points that we call $W_\pm$, i.e., $\varphi_{W_-}=\pi/2$ and $\varphi_{W_+}=0$. The corresponding radial variables are found to be
\begin{equation}
R_{W_\pm}=\sqrt{\frac{1}{2\beta}\left(1\mp\frac{2}{\tilde h}\right)} \,\, .
\label{eq:radialcoordinatesingleoscillator}
\end{equation}
Instead, for $\varphi\neq0,\pi/2$, one has from Eq.~\eqref{eq:multiplescalepolarcoordinatesdynamic1p}
\begin{equation}
R^2=\frac{1}{\beta}\left(\frac{1}{2}-\frac{1}{\tilde h}\right)\frac{1}{\cos^2(\varphi)+3\,\sin^2(\varphi)} \,\, ,
\end{equation}
and if this is used in Eq.~\eqref{eq:multiplescalepolarcoordinatesdynamic2p}, one has the solution for the angular variable
\begin{equation}
\cos(\varphi)=\pm\sqrt{\frac{1}{2}\left(1+\frac{4}{\tilde h}\right)}\,\, ,
\label{eq:angularvariableequation}
\end{equation}
for $\tilde h>4$. Equation~\eqref{eq:angularvariableequation} identifies two additional fixed points that we call $P$ and $Q$, whose angular and radial variables are therefore
\begin{subequations}
\begin{align}
\varphi_P&={\rm arccos}\left[\sqrt{\frac{1}{2}\left(1+\frac{4}{\tilde h}\right)}\,\right]\\
\varphi_Q&=\pi-{\rm arccos}\left[\sqrt{\frac{1}{2}\left(1+\frac{4}{\tilde h}\right)}\,\right]\\
R_{P,Q}&=\frac{1}{2\sqrt{\beta}} \,\, .
\end{align}
\end{subequations}
We therefore have the following picture: for $\beta>0$, $W_-$ is found for any $\tilde h>0$, $W_+$ is found for $\tilde h>2$, whereas $P$ and $Q$ are found for $\tilde h>4$. From the expression of $\mathbf{J}_1$ in Appendix~\ref{sec:expressionforthejacobianmatrix}, the eigenvalues of the Jacobian are found to be
\begin{equation}
\hspace{-0.2cm}\lambda_\pm=-\frac{\omega_0}{2}\pm\frac{\tilde h\omega_0}{2}\sqrt{\frac{1}{4}-3\beta\,R^2+9\,\beta^2R^4\,\cos^2(2\varphi)} \,\, ,
\label{eq:eigenvaluesjacobianpumpdepletion}
\end{equation}
for a given fixed point, i.e., for the specific values of $R$ and $\varphi$. The eigenvalues of the Jacobian matrix are independent of $\beta$, since $R\sim1/\sqrt{\beta}$ for all fixed points with $R\neq0$.

There are four main different situations depending on the values of the system parameters that one can consider: first, for $\beta=0$ and $\tilde h<2$, the origin is the only fixed point of the system and it is a stable node [Fig.~\ref{fig:flowequationaphi0}, panel (a)], which becomes a saddle point when $\tilde h>2$ (not shown).

Second, for $\beta>0$ and $\tilde h<2$ [Fig.~\ref{fig:flowequationaphi0}, panel (b)], two additional fixed points ($W_-$) appear in addition to the origin. We see that, at the origin ($R=0$) the eigenvalues of the Jacobian are $\lambda_{\pm;O}=-\omega_0/2\pm\omega_0\tilde h/4$, and therefore they are both real and negative if $\tilde h<2$, whereas the eigenvalues of the Jacobian for the $W_-$ points are $\lambda_{\pm;W_-}=-\omega_0/2\pm(\omega_0+3\tilde h)/2$, which are always real and with opposite sign. The points $W_-$ are always saddle points, and therefore in the case $\tilde h<2$ only the origin is a stable point also for $\beta>0$.

Third, for $2<\tilde h<4$, two stable nodes ($W_+$) are born in pairs from the origin via a saddle-node bifurcation, after which the origin becomes a saddle point, independent of $\beta$ [Fig.~\ref{fig:flowequationaphi0}, panel (c)]. This can be seen by looking at the eigenvalues of the Jacobian matrix: for the fixed points $W_+$ and $W_-$, the eigenvalues of the Jacobian are $\lambda_{\pm;W_+}=-\omega_0/2\pm\omega_0|3-\tilde h|/2$ and $\lambda_{\pm;W_-}=-\omega_0/2\pm\omega_0(3+\tilde h)/2$. In this range of $\tilde h$, the point $W_-$ is always a saddle point, whereas the point $W_+$ is a stable node, with both eigenvalues of the Jacobian real and negative, whereas the origin (whose eigenvalues of the Jacobian matrix are $\lambda_{\pm;O}=-\omega_0/2\pm\omega_0\tilde h/4$, see above), is a saddle points when $\tilde h>2$. In this situation, for the trajectories flowing to the fixed point $W_+$, the imaginary part of the complex amplitude $A_I$ is suppressed ($A_I\rightarrow0$), and the real part $A_R$ is stabilized to some nonzero value. This situations corresponds to squeezing.

Fourth, for $\tilde h>4$, two new stable attractors ($P$ and $Q$) are born from $W_+$ via a saddle-node bifurcation, after which the fixed point $W_+$ becomes a saddle point [Fig.~\ref{fig:flowequationaphi0}, panel (d)]. For the fixed points $Q$ and $P$, the eigenvalues of the Jacobian matrix are $\lambda_{\pm;Q,P}=-\omega_0/2\pm(\omega_0/2)\sqrt{9-h^2/2}$. The fixed points $P$ and $Q$ are therefore stable nodes with both eigenvalues real and negative for $4<\tilde h<3\sqrt{2}$, and they are stable focuses (with eigenvalues with negative real part and nonzero imaginary part) for $\tilde h>3\sqrt{2}$.

One can see that the effect of having $\phi>0$ is to rigidly rotate the flow of the nonlinear equation by an angle of $\phi/2$. Therefore, in computing the position of the fixed points, one simply has to redefine the angles as $\varphi_{W_\pm}\rightarrow\varphi_{W_\pm}+\phi/2$ and $\varphi_{P,Q}\rightarrow\varphi_{P,Q}+\phi/2$, while the radial coordinates and the eigenvalues of the Jacobian matrix remain unaffected by $\phi$.

\subsubsection{Kerr nonlinearity}
\label{sec:decoupledparametricoscillators2}
We here recall the stability diagram of the nonlinear Mathieu's equation with Kerr nonlinearity in the resonant case (see for instance also Ref.~\cite{10.1143/PTP.98.755}). Focusing on the case of $\phi=0$, the nonlinear equations for the slow-varying amplitude are ($\tilde\tau=\omega_0\tau$)
\begin{subequations}
\begin{align}
&\frac{\partial A_R}{\partial\tilde\tau}=\left(\frac{\tilde h}{4}-\frac{1}{2}\right)A_R+\frac{3\beta}{2}\,{|A|}^2\,A_I\\
&\frac{\partial A_I}{\partial\tilde\tau}=\left(-\frac{\tilde h}{4}-\frac{1}{2}\right)A_I-\frac{3\beta}{2}\,{|A|}^2\,A_R\,\, .
\end{align}
\label{eq:nonlinearmathieuequationflow22}
\end{subequations}
From Eq.~\eqref{eq:nonlinearmathieuequationflow22}, one can see that there are two possible situations: when $\beta=0$, the origin is a stable node for $\tilde h<2$ and it is a saddle point for $\tilde h>2$, as in the case of Appendix~\ref{sec:decoupledparametricoscillators}. For $\beta>0$, the origin is the only fixed point for $\tilde h<2$ and it is a stable point. For $\tilde h>2$, the origin becomes a saddle node and two additional stable fixed points (which we denote by $P$) are born from the origin via a saddle-node bifurcation. As done in Appendix~\ref{sec:decoupledparametricoscillators}, we express the coordinates of the fixed point $P$ in polar coordinates, whose radial coordinate is
\begin{equation}
R_P=\frac{1}{\sqrt{3\beta}}{\left(\frac{\tilde h^2}{4}-1\right)}^{1/4} \,\, ,
\label{eq:fixedpointkerr}
\end{equation}
and the angular coordinate is
\begin{equation}
\varphi_P=-{\rm arctan}\left(\,\sqrt{\frac{\tilde h-2}{\tilde h+2}}\,\right) \,\, .
\end{equation}
The eigenvalues of the Jacobian are found to be
\begin{equation}
\lambda_{\pm}=-\frac{\omega_0}{2}\pm\frac{\omega_0}{2}\sqrt{\frac{\tilde h^2}{4}+3\,\tilde h\beta\,R^2\,\sin(2\varphi)-27\,\beta^2\,R^4} \,\, ,
\end{equation}
for a given fixed point. As in Eq.~\eqref{eq:eigenvaluesjacobianpumpdepletion}, the eigenvalues of the Jacobian are independent of $\beta$. For the origin ($R=0$), one has as in Appendix~\ref{sec:decoupledparametricoscillators} the eigenvalues $\lambda_{\pm;O}=-\omega_0/2\pm\omega_0\tilde h/4$. Instead, for the point $P$, one has $\lambda_{\pm;P}=-\omega_0/2\pm(\omega_0/2)\sqrt{5-\tilde h^2}$. The fixed points $P$ are the only fixed points in addition to the origin (in the resonant case $\gamma=2\omega_0$) that are found for all $\tilde h>2$, and they are stable nodes for $2<\tilde h<\sqrt{5}$ and stable focuses for $\tilde h>\sqrt{5}$. An example of the flow is shown in Fig.~\ref{fig:flowequationaphi1}, in the case of $\beta>0$ and for the below-threshold case $\tilde h<2$ [panel (a)], and for the above-threshold case $\tilde h>2$ [panel (b)].

\subsection{Oscillation threshold and beats}
\label{sec:oscilaltionthresholdandbeats}
In this appendix, we discuss the stability properties of the origin as a function of the system parameters for the system discussed in Sec.~\ref{sec:nonlinearcasebymultiplescaleanalysis}, in the presence of both energy-preserving and dissipative coupling (see Sec.~\ref{sec:dissipativecouplngandcim}). The results that we discuss in this section are valid also for $\beta=0$, since the stability properties of the origin are unaffected by the nonlinearity.

The informations regarding the position of the critical line for the bifurcation separating the region of the phase diagram in which the origin is a stable attractor (below threshold) from the one in which the origin is unstable (limit cycle, CIM or synchronization region) can be determined by studying the Jacobian matrix at the origin, i.e., $A=B=0$ (see Appendix~\ref{sec:expressionforthejacobianmatrix}). The eigenvalues of the Jacobian matrix at the origin are found to be
\begin{eqnarray}
&&\lambda_{\pm,\pm}=-\frac{\omega_0}{2}\nonumber\\
&&\pm\frac{\omega_0}{4}\sqrt{\tilde h^2-4\,(\tilde r^2-\tilde\alpha^2)\pm i\,4\,\tilde h\,\sqrt{\tilde r^2-\tilde\alpha^2}\cos\left(\frac{\phi}{2}\right)} \,\, . \nonumber\\
\label{eq:mathieumultiplescalenalaysis522cphigfixed2}
\end{eqnarray}
The origin is \rev{a stable point} if all the eigenvalues in Eq.~\eqref{eq:mathieumultiplescalenalaysis522cphigfixed2} have negative real part. \rev{In this case, the largest negative real part gives the decay rate to the trivial solution $A=B=0$ below the oscillations threshold: $\tau^{-1}_{\rm decay}=-g\,{\rm max}\{{\rm Re}[\lambda_{\pm,\pm}]\}=g\,{\rm min}|{\rm Re}[\lambda_{\pm,\pm}]|$. For sufficiently long times, below threshold, the decay to the trivial solution is then $x_1(t),x_2(t)\sim e^{-t/\tau_{\rm decay}}$}. We can therefore have two regimes:

\vspace{0.2cm}
\textit{\underline{(i) Case $\tilde r>\tilde\alpha$ - Beating regime}}. By defining for convenience the function
\begin{eqnarray}
&&Y(\tilde h,\tilde r,\tilde \alpha,\phi)=\tilde h^2-4\,\left(\tilde r^2-\tilde\alpha^2\right)\nonumber\\
&&+\sqrt{{\left[\tilde h^2-4\left(\tilde r^2-\tilde\alpha^2\right)\right]}^2+16\,\tilde h^2\left(\tilde r^2-\tilde\alpha^2\right)\,\cos^2\left(\frac{\phi}{2}\right)} \,\, ,\nonumber\\
\label{eq:equationforthefunctionright2dissipativecoupling}
\end{eqnarray}
one can see that, in the $\tilde h$ vs. $\tilde r$ plane, the origin is stable point if $\tilde h<\tilde h_{\rm th}(\tilde r,\tilde\alpha,\phi)$, where $\tilde h_{\rm th}(\tilde r,\tilde\alpha,\phi)$ is identified by the contour line $Y(\tilde h, \tilde r,\tilde\alpha,\phi)=8$, which yields
\begin{equation}
\tilde h_{\rm th}(\tilde r,\tilde\alpha,\phi)=2\sqrt{\frac{\left(\tilde r^2-\tilde\alpha^2\right)+1}{\left(\tilde r^2-\tilde\alpha^2\right)\,\cos^2(\phi/2)+1}} \,\, .
\label{eq:boundaryoscillationthreshold}
\end{equation}
Notice that, for generic $\phi$, one has from Eq.~\eqref{eq:boundaryoscillationthreshold} $\tilde h_{\rm th}(\tilde r,\tilde\alpha,\phi)\geq2$, for all $\tilde r$, where the lower bound $\tilde h_{\rm th}=2$ is found for $\phi=0$, and it is independent of $\tilde r$, which is correctly the threshold condition \rev{$h/(2g)=1$} discussed in Eq.~\eqref{eq:parametricoscillatorsolutionfloquet} at the parametric resonance ($\epsilon=0$). For a generic $\phi$, the threshold for parametric oscillations depends on the strength of the coupling.

The imaginary part of the eigenvalues in Eq.~\eqref{eq:mathieumultiplescalenalaysis522cphigfixed2}, when present, is what determines the presence or absence of beats. From Eq.~\eqref{eq:mathieumultiplescalenalaysis522cphigfixed2}, one can see that
\begin{equation}
\left|{\rm Im}[\lambda_{\pm,\pm}]\right|=\frac{\sqrt{\tilde r^2-\tilde\alpha^2}}{2}\sqrt{\frac{(\tilde r^2-\tilde\alpha^2+1)\cos^2(\phi/2)}{(\tilde r^2-\tilde\alpha^2)\cos^2(\phi/2)+1}} \,\, ,
\label{eq:imaginaryparteigenvalues}
\end{equation}
and therefore the angular frequency of the limit cycle at threshold is $\omega_{\rm LC,th}(\tilde r,\tilde\alpha,\phi)=\omega_0g\,\left|{\rm Im}[\lambda_{\pm,\pm}]\right|$.

It is now interesting to compare the results in Eq.~\eqref{eq:boundaryoscillationthreshold} and Eq.~\eqref{eq:imaginaryparteigenvalues} with the one discussed in the linear case in Sec.~\ref{sec:instabilityregions}. One can see that, above threshold for $\phi<\pi$, the origin is always an unstable point with $\left|{\rm Im}[\lambda_{\pm,\pm}]\right|>0$, and therefore the oscillators always display beats. In such situation, for $\beta=0$, no limit cycle can stabilize the amplitude of the beats and therefore the long-time dynamics of the oscillators is given by an exponential amplification with the beats superimposed. This situation corresponds to the one discussed in the linear case in Sec.~\ref{sec:instabilityregions} by means of Floquet theorem, in the case in which the central instability region was present [see for example Fig.~\ref{fig:instabilityregions}, panels (a) and (b)].

Instead, at $\phi=\pi$ and above the threshold identified by $\tilde h_{\rm th}(\tilde r)=2\sqrt{\tilde r^2+1}$, one has $\left|{\rm Im}[\lambda_{\pm,\pm}]\right|=0$. In this case, the origin is unstable and amplification occurs without beats. This situation, for $\beta=0$, corresponds to the situation shown in Fig.~\ref{fig:instabilityregions}, panel (c), in particular when the system is in the region in which the two outer instability regions overlap (around $\gamma=2\Omega_r\simeq2\omega_0$). On top of these behaviours, found also in the linear case, the interplay between $\beta,g>0$ is what eventually stabilizes the beats with the presence of the limit cycle (for $\phi<\pi$) or the synchronization with the presence of additional stable attractors (for $\phi=\pi$) as the oscillation threshold [Eq.~\eqref{eq:boundaryoscillationthreshold}] is crossed.

\vspace{0.2cm}
\textit{\underline{(ii) Case $\tilde r<\tilde\alpha$ - CIM regime}}. In this case, in the regime of interest, the eigenvalues in Eq.~\eqref{eq:mathieumultiplescalenalaysis522cphigfixed2} are real. This first implies that, even when the origin becomes unstable, no limit cycle is found. The origin is an unstable point when $\lambda_{+,+}>0$, and therefore the threshold is identified by the condition
\begin{equation}
\tilde h^2+4\,\tilde h\,\sqrt{(\tilde\alpha^2-\tilde r^2)}\,\cos\left(\frac{\phi}{2}\right)+4\left[(\tilde\alpha^2-\tilde r^2)-1\right]=0 \,\, ,
\end{equation}
from which for the requirement $\tilde h\in\mathbb{R}$ one obtains
\begin{eqnarray}
&&\tilde h_{\rm th}(\tilde r,\tilde\alpha,\phi)=2\sqrt{1+(\tilde\alpha^2-\tilde r^2)\left[\cos^2\left(\frac{\phi}{2}\right)-1\right]}\nonumber\\
&&\hspace{2cm}-2\sqrt{\tilde\alpha^2-\tilde r^2}\,\cos\left(\frac{\phi}{2}\right)\,\, .
\end{eqnarray}
As discussed in Sec.~\ref{sec:dissipativecouplngandcim}, above this threshold, the region with two stable fixed points is found.

\subsection{Alternative form of the coupling - Multiple-scale analysis}
\label{sec:inphasecouplingmultiplescaleanalysis}
We report in this appendix the result of the calculation of the multiple-scale equations, using an alternative and commonly-used form of the coupling, and $\alpha=0$. We write the equation of motion [Eq.~\eqref{eq:coupledparametricoscillatorequationswithnonlinearitydynamicalphigfixed}] as
\begin{eqnarray}
&&\hspace{-0.5cm}\ddot x_1+\omega^2_0\left[1+h\left(1-\beta\,x^2_1\right)\,\sin(\gamma t)\right]x_1 \nonumber\\
&&\hspace{2.5cm}+\omega_0g\,\dot x_1+\omega_0^2r\, x_2=0\,\,\,
\label{eq:coupledparametricoscillatorequationswithnonlinearitydynamicalphigfixedphase1}
\end{eqnarray}
\begin{eqnarray}
&&\hspace{-0.5cm}\ddot x_2+\omega^2_0\left[1+h\left(1-\beta\,x^2_2\right)\,\sin(\gamma t+\phi)\right]x_2 \nonumber\\
&&\hspace{2.5cm}+\omega_0g\,\dot x_2+\omega_0^2r\, x_1=0 \,\, .
\label{eq:coupledparametricoscillatorequationswithnonlinearitydynamicalphigfixedphase2}
\end{eqnarray}
The multiple-scale equations governing the dynamics of the slow-varying amplitudes of $x_1$ and $x_2$ in Eqs.~\eqref{eq:coupledparametricoscillatorequationswithnonlinearitydynamicalphigfixedphase1} and~\eqref{eq:coupledparametricoscillatorequationswithnonlinearitydynamicalphigfixedphase2} are obtained as done for Eqs.~\eqref{eq:nonlinearmathieuequationflow2220coupledphihfixed}. One obtains ($\tilde\tau=\omega_0\tau$)
\begin{widetext}
\begin{subequations}
\begin{align}
&\frac{\partial A_R}{\partial\tilde\tau}=\left[\frac{\tilde h}{4}-\frac{1}{2}-\frac{\beta\tilde h}{2}\left(A^2_R+3\,A^2_I\right)\right]A_R-\frac{\tilde r}{2}\,B_I \qquad \frac{\partial A_I}{\partial\tilde\tau}=\left[-\frac{\tilde h}{4}-\frac{1}{2}+\frac{\beta\tilde h}{2}\left(A^2_I+3\,A^2_R\right)\right]A_I+\frac{\tilde r}{2}\,B_R\\
&\frac{\partial B_R}{\partial\tilde\tau}=\frac{\tilde h}{4}\left[B_R\,\cos(\phi)+B_I\,\sin(\phi)\right]-\frac{1}{2}\,B_R-\frac{\beta\tilde h}{2}\left[B^3_R\,\cos(\phi)+2\,B^3_I\sin(\phi)+3\,B_RB^2_I\,\cos(\phi)\right]-\frac{\tilde r}{2}\,A_I\\
&\frac{\partial B_I}{\partial\tilde\tau}=\frac{\tilde h}{4}\left[B_R\,\sin(\phi)-B_I\,\cos(\phi)\right]-\frac{1}{2}\,B_I-\frac{\beta\tilde h}{2}\left[2B^3_R\,\sin(\phi)-B^3_I\,\cos(\phi)-3\,B^2_RB_I\,\cos(\phi)\right]+\frac{\tilde r}{2}\,A_R \,\, .
\end{align}
\label{eq:nonlinearmathieuequationflow2220coupledphihfixed22}
\end{subequations}
\end{widetext}
By comparing Eqs.~\eqref{eq:nonlinearmathieuequationflow2220coupledphihfixed22} with Eqs.~\eqref{eq:nonlinearmathieuequationflow2220coupledphihfixed}, one can verify that the two set of equations describe the same dynamics if we perform in Eq.~\eqref{eq:nonlinearmathieuequationflow2220coupledphihfixed} the rotation $B_R\rightarrow-B_I$, $B_I\rightarrow B_R$ and redefine $\phi\rightarrow\pi+\phi$. This proves that, in the limit of small coupling, the long-time dynamics of the model with the coupling as in Eq.~\eqref{eq:coupledparametricoscillatorequationswithnonlinearitydynamicalphigfixed} and pump dephasing $\phi$ is equivalent to the long-time dynamics of the model with the coupling as in Eqs.~\eqref{eq:coupledparametricoscillatorequationswithnonlinearitydynamicalphigfixedphase1} and~\eqref{eq:coupledparametricoscillatorequationswithnonlinearitydynamicalphigfixedphase2} with pump dephasing $\pi+\phi$.

\subsection{Expression of the Jacobian matrix of the system in Eq.~\eqref{eq:nonlinearmathieuequationflow2220coupledphihfixed}}
\label{sec:expressionforthejacobianmatrix}
We here explicitly report for completeness the expression of the Jacobian matrix computed around a given point $A_R,A_I,B_R,B_I$ from the system in Eq.~\eqref{eq:nonlinearmathieuequationflow2220coupledphihfixed}. The Jacobian matrix can be compactly written as
\begin{equation}
\mathbf{J}(A_R,A_I,B_R,B_I)=\omega_0\left(\begin{array}{cc}\mathbf{J}_1&\mathbf{J}_2\\\mathbf{J}_3&\mathbf{J}_4(\phi)\end{array}\right) \,\, ,
\end{equation}
where we define the $2\times2$ blocks as follows: \rev{one block for the oscillator A}
\begin{equation}
\mathbf{J}_1=\left(\begin{array}{cc}
\cfrac{\tilde h}{4}-\cfrac{1}{2}-\cfrac{3\beta\tilde h}{2}\,{|A|}^2 \,\,&\,\, -3\beta\tilde h\,A_RA_I\\\\
3\beta\tilde h\,A_RA_I \,\, & \,\, -\cfrac{\tilde h}{4}-\cfrac{1}{2}+\cfrac{3\beta\tilde h}{2}\,{|A|}^2
\end{array}\right)
\end{equation}
\rev{one} phase-dependent block \rev{for the oscillator $B$}
\begin{widetext}
\begin{equation}
\mathbf{J}_4(\phi)=\left(\begin{array}{cc}
\cfrac{\tilde h\,\cos(\phi)}{4}-\cfrac{1}{2}-\cfrac{3\beta\tilde h\,\cos(\phi)}{2}\,{|B|}^2 \,\,&\,\, \cfrac{\tilde h\,\sin(\phi)}{4}-3\beta\tilde h\,\left[B^2_I\,\sin(\phi)+B_RB_I\,\cos(\phi)\right]\\\\
-\cfrac{\tilde h\,\sin(\phi)}{4}-3\beta\tilde h\,\left[B^2_R\,\sin(\phi)-B_RB_I\,\cos(\phi)\right] \,\, & \,\, -\cfrac{\tilde h\,\cos(\phi)}{4}-\cfrac{1}{2}+\cfrac{3\beta\tilde h\,\cos(\phi)}{2}\,{|B|}^2
\end{array}\right) \,\, .
\end{equation}
\end{widetext}
\rev{and eventually the block describing the coupling between the oscillator $A$ and the oscillator $B$}
\begin{equation}
\mathbf{J}_2=-\mathbf{J_3}=\left(\begin{array}{cc}\cfrac{\tilde r}{2}\,\, &\,\, 0\\0\,\, &\,\, \cfrac{\tilde r}{2}\end{array}\right) \,\, ,
\end{equation}


%

\end{document}